\documentstyle[preprint,eqsecnum,aps,epsf]{revtex}
\newcommand{\be}{\begin{equation}}
\newcommand{\ee}{\end{equation}}
\newcommand{\ba}{\begin{eqnarray}}
\newcommand{\ea}{\end{eqnarray}}
\newcommand{\robold}{\mbox{\boldmath$\rho$}}
\begin{document}
\title{
${}^{~~~~~~~~~~~~~~~~~~~~~~~~~{published ~in ~the}}$\\
${}^{\it{Astrophysical~Journal, ~Vol. ~575, ~No~2, ~pp.~886~-~899~(2002)}}$\\
${}^{}$\\
${}^{}$\\Mechanical Alignment of Suprathermal Paramagnetic Cosmic-Dust 
Granules: the  Cross-Section Mechanism\\}
\author{Michael Efroimsky\\}
\address{Institute for Mathematics and Its Applications,
University of Minnesota,\\ 207 Church Street SE, Suite 400, Minneapolis
MN 55455\\}

\maketitle

\begin{abstract}
We develop a comprehensive quantitative description of the cross-section
mechanism discovered several years ago by Lazarian. 
This is one of the processes that determine grain orientation in 
clouds of suprathermal cosmic dust. The cross-section mechanism
manifests itself when an ensemble of suprathermal 
paramagnetic granules is placed in a magnetic field and
is subject to ultrasonic gas bombardment. The mechanism yields dust 
alignment whose efficiency depends upon two factors: the geometric 
shape of the granules, and the angle $\,\Phi\,$ between the magnetic line and 
the gas flow. We calculate the quantitative measure of this alignment, and study
its dependence upon the said factors. It turns out that, irrelevant of the
grain shape, the action of a flux does not lead to alignment if
$\,\,\Phi\,=\,\arccos (1/\sqrt{3})$.

\end{abstract}


\section{Introduction. The Physical Nature of Effect.}

Polarisation of the starlight is a long-known effect. Due to the 
correlation of polarisation with reddening, this phenomenon is put 
down to the orientation of particles in the cosmic-dust nebulae (Hall
1949, Hiltner 1949). This orientation causes differential extinction of
electromagnetic waves of different polarisations (effect known in
optics as linear dichroism), and provides a remarkable example of
order emerging in a seemingly chaotic system. 

In a nutshell, the polarization is due to the fact 
that the grains are non-spherical (i.e., have different cross sections in the 
body frame) and these non-spherical cross sections are somehow aligned
within the cloud.
A remarkable feature of this  phenomenon is that, whatever orientational
mechanisms show themselves in the dust-grain rotational dynamics, the 
orientation always takes place relative to the interstellar magnetic field.

Thus, whenever the word ``orientation'' is used, it always is an euphemism for
``orientation with respect to the magnetic line''. Another semantic issue 
is the conventional difference between the meaning of words 
``orientation'' and ``alignment''. Historically, the word ``orientation''
has been used to denote existence of {\it one} preferred direction in a 
physical setting. The analysis of cosmic-dust orientation has demonstrated that
in all known cases this preferred direction is equivalent to its
opposite. Simply speaking, were an instantaneous inversion of the magnetic
field in the dust cloud possible, it would not alter the
orientation of the dust ensamble. Because of such an invariance, 
the word ``orientation'' is often avoided and substituted by the
word ``alignment'' which is thereby imparted with the desired broader 
meaning. (The adjective ``orientational'', though, remains in
use.) We shall abide by this verbal code.

The rotational dynamics of an interstellar particle is determined by a 
whole bunch of accompanying physical processes whose combination 
produces a variety of orientational mechanisms. Which of these come into play in a particular physical setting, depends
upon the suprathermality of the dust cloud. Suprathermal are, by 
definition, grains which spin so rapidly that their averaged (over 
the dust ensemble) rotational kinetic energy $\,<E_{\small{rot}}>\,$ 
much exceeds the (multiplied 
by the Bolzmann constant $\kappa$) temperature $T_{\small{gas}}$ of 
the surrounding environment. The suprathermality degree is then 
introduced as the following ratio:
\be
\beta\;=\;<E_{\small{rot}}>/{\kappa}T_{\small{gas}}\;\;\;\;.\;\;
\label{1.1}
\ee
Dust ensembles with $\,\beta\,$ not much different from unity are
called thermal or Brownian. Clouds with $\,\beta\,\gg\,1$ are called 
suprathermal. In the observable Universe, values of $\,\beta\,$ of
order $\,10^2\,$ are not unusual.

The leading reason for suprathermal rotation is formation of
$\bf{H_2}$ molecules at the defects on the granule surface: over such a 
defect (called active site), two atoms of $\bf{H}$ couple to form a
molecule, ejection whereof applies an uncompensated torque to the
granule surface (Purcell 1979). These, so-called spin-up torques keep 
emerging at each active site until the site gets ``poisoned'' through the
everlasting accretion. After that, some other active site will
dominate the spin dynamics of the grain, by its $\bf{H_2}$
``rocket''. This change of spin state will, with some probability, go 
through a short-term decrease, to thermal values, of the grain's
angular velocity. Such breaks are called ``cross-overs'' or ''flip-overs''. 

Another pivotal circumstance to be mentioned here is the existing evidence
of paramagnetic nature of a considerable portion of dust particles (Whittet
1992) which makes them subject to the Barnett effect. This phenomenon
takes place in para-~~and ferromagnetics, due to interaction between
the spins of unpaired electrons and the macroscopic rotation of
crystal lattice (Stoner 1934). This coupling has its origin in the 
angle-dependent terms in the dipole-dipole interaction of neighbouring 
spins. It spontaneously endows a rotating para- or ferromagnetic body
with a magnetic moment parallel to the angular
velocity\footnote{Another contribution to the magnetisation comes from
the electric charge carried by the granule.} (Lazarian \& Roberge
1997). Purcell offered the following illustrative
explanation of the effect. If a rotating body contains equal amount 
of spin-up and spin-down unpaired electrons, its magnetisation is
nil. Its kinetic energy would decrease, with the total angular
momentum remaining unaltered, if some share of the entire 
angular momentum could be transferred to the spins by turning some of
the unpaired spins over (and, thus, by dissipating some energy). This 
potential possibility is brought to pass through the said coupling. 

An immediate outcome from granule magnetisation is the subsequent coupling of 
the magnetic moment $\;\bf M\;$ with the interstellar magnetic field
$\;\bf B\;$: the magnetic
moment precesses about the magnetic line. What is important, is that 
this precession goes at an intermediate rate. On the one hand, it
is slower than the grain's spin about its instantaneous rotation axis. On the
other hand, the precession period is much shorter than the typical time 
scale at which the relative-to-$\bf B\;$ alignment gets established\footnote{
A more exact statement, needed in Section V below, is that the period
of precession (about $\;\bf B\;$) of the magnetic moment $\;\bf M\;$ (and of
$\,Z\,$ aligned therewith) is much shorter than the mean time between two
sequent flip-overs of a spinning granule (Purcell 1979, Roberge et al 1993)}. 
The latter was proven by Dolginov \& Mytrofanov (1976), for magnetisation
resulting from the Barnett effect, and by Martin (1971), for magnetisation 
resulting from the grain's charge.

If we disembody the core idea of the Barnett effect from its
particular implementation, we shall see that it is of quite a general
sort: a free rotator, though conserves its angular momentum, tends
to minimise its kinatic energy through some dissipation
mechanism(s). This fact, neglected in the Euler-Jacobi theory of
unsupported top, makes their theory inapplicable at time
scales comparable to the typical time of dissipation (Efroimsky 2002). 
The needed generalisation of the insupported-top dynamics constitutes 
a mathematically involved area of study (Efroimsky 2000), which provides  
ramifications upon wobbling asteroids and comets (Efroimsky 2001), 
rotating spacecraft, and even precessing neutron stars (Tr\"{u}mper et al
1986, Alpar \& \"{O}gelman 1987, Bisnovatyi-Kogan \& Kahabka 1993,
Stairs et al 2000). Fortunately, 
in the case of cosmic-dust physics, we need only some basics of this
theory. A free 
rotator has its kinetic energy minimised (with its angular momentum being
fixed) when the rotation axis coincides with the axis of major
inertia. In this, so-called principal state, the major-inertia axis $\,Z\,$,
the angular velocity $\;\bf \Omega\,$, and the angular-momentum vector 
$\;\bf J\;$ are all aligned. In other, complex rotation states, both 
the maximal-inertia axis $\,Z\,$ and angular-velocity $\;\bf \Omega
\,$ precess about the angular momentum $\;\bf J\;$. This precession is also 
called ``wobble'' or ``tumbling'', in order to distinguish it from the 
precession of the magnetic moment $\;\bf M\;$ about the 
magnetic line. Similarly, the wobble relaxation (i.e., gradual alignment of
axis $\,Z\,$ and of $\;\bf \Omega\,$ toward $\;\bf J\;$) should 
be distinguished from the granule alignment relative to the magnetic 
field $\;\bf B\;$: the latter effect is the eventual target of our treatise, 
while the former is merely a trend in the tapestry. Still, the wobble
relaxation is far more than a mere technicality: it is important to know if a 
typical time of the wobble relaxation 
is much less than the typical times of the external interactions (like, say, 
the period of precession of $\;\bf M\;$ about $\;\bf B\;$).
In case the wobble relaxation is that swift, one may assume that the
precession (about $\;\bf B\;$) of the magnetic moment $\;\bf M\;$ is the 
same as precession of the angular momentum $\;\bf J\;$ about $\;\bf
B\;$: indeed, in this case, both $\;\bf J\;$ and the major-inertia
axis $\,Z\,$ will be aligned with $\;\bf \Omega\,$ which is parallel to 
$\;\bf M\;$. 
This parallellism of all four vectors is often called not alignment
but ``coupling'', to distinguish it from the alignment relative to 
$\;\bf B\;$. This coupling is enforced by two different processes. One
is an effect kin to that of Barnett: tumbling of $\;\bf \Omega\;$
relative to conserved\footnote{At time scales shorter than a precession cycle 
about $\;\bf B\;$.} $\;\bf J\;$ (and, therefore, relative to an 
inertial observer) yields periodic remagnetisation of the material,
which results in dissipation. The other effect is the anelastic 
dissipation: in a complex rotational state the points inside the 
body experience time-dependent accelaration that produces alternate 
stresses and strains. Anelastic phenomena entail inner friction 
(which may be understood also in terms of a time lag between the
strain and stress). The contributions from the Barnett and anelastic 
effects to the coupling were compared by Purcell in his
long-standing cornerstone work (Purcell 1979). Purcell came to an 
unexpected conclusion that the input from the Barnett effect much outweights 
that from anelasticity. Having set out the calculations, Purcell continued with
phrase: ``It may seem surprising that an effect as feeble, by most standards of
measure, as the Barnett effect could so decisively dominate the grain 
dynamics''. After that, Purcell tried to pile up some qualitative evidence, to
buttress up the unusual result. Still, the afore quoted emotional passage 
reveals that, most probably, Purcell's tremendous physical intuition 
signalled him that something had been overlooked in his study.  
An accurate treatment (Lazarian \& Efroimsky 1999) shows that the 
anelastic dissipation is several orders of magnitude more effective 
than presumed, and in many physical settings it dominates over the 
Barnett dissipation. The case of suprathermal dust is one such setting.
Without going into redundant details, we would mention that 
combination of the two dissipation processes provides at least partial
alignment of $\;Z\;$ and $\;\bf \Omega\;$ toward $\;\bf J\;$ in
Brownian clouds, and it provides a perfect alignment in suprathermal ones.

The presently known mechanisms of grain alignment can be classified into three 
cathegories: mechanical mechanisms, paramagnetic mechanisms, and via
radiative torques. The latter mechanism was addressed in (Dolginov \&
Mytrofanov (1976); Lazarian 1995b; Draine and Weingartner 1996, 1997). It has 
not yet been well understood. The 
paramagnetic alignment is due to the Davis-Greenstein (1951) 
mechanism (initially suggested for 
Brownian dust particles), and due to the Purcell
(1979) mechanism (which is a generalisation of the Davis-Greenstein
mechanism, to the suprathermal case). The Davis-Greenstein
and Purcell processes operate to bring the granule's rotation
axis (which is, as explained above, fully or partially aligned with
the granule's major-inertia axis), into 
parallelism with the magnetic line. This happens because 
precession of grain's spin axis about $\;\bf B\;$ entails 
material remagnetisation\footnote{It is
assumed that the grain is either paramegnetic (Davis \& Greenstein
1951) or ferromagnetic (Spitzer \& Tukey 1951), (Jones \& Spitzer
1967). The case of a diamagnetic granule has not had been addressed in
the literature so far.} and, therefore, dissipation resulting in a 
slow removal of the rotation component orthogonal to $\bf B$. The
induced alternating magnetisation $\bf M$ will lag behind 
rotating $\bf B$, giving birth to a nonzero torque equal, in the body
frame, to 
${\bf{M}}\times{\bf{B}}$. It can be shown (Davis \& Greenstein 1951) 
that this torque will entail steady decrease of the orthogonal-to-$\bf 
B$ component of the angular velocity\footnote{A rigorous analysis of the 
Davis-Greenstein process should be carried out in the language of 
Fokker-Planck equation (Jones \& Spitzer 1967).}. The 
so-called\footnote{The word 
``so-called'' is very much in order here, because the mechanical
mechanisms, too, provide alignment relative to the magnetic line, and
their very name, ``mechanical'' simply reflects the fact that these
effects are not purely magnetic but involve the grains' mechanical 
interaction with the interstellar wind.} mechanical alignment
comprises the Gold (1952) mechanism,  and those of Lazarian 
(1995 a,b,c,d,e). Lazarian suggested two mechanisms: the cross-over one 
and the cross-section one, and they show themselves in the case of
suprathermal grains only.

The nature of the Gold mechanism is the following. Each collision of the dust
particle with an atom or a molecule of the streaming gas adds to the
particle's angular momentum a portion perpendicular to the relative 
velocity. As explained in one of the above footnotes, the
major-inertia axis of the body tends to align with the angular 
momentum. One, hence, may say that the interstellar wind will spin-up 
the granule so that its
maximal-inertia axis will ``prefer'' positions perpendicular to the
wind. Since the said major inertia axis is, roughly, the shortest
dimension of the rotator, one may deduce that, statistically, the
particles tend to rotate with their shortest axes orthogonal to the 
gas flow. This picture is, though, complified by the precession of the 
magnetic moments (and of the angular momenta that tend, for the afore
mentioned reason, to align with the magnetic moments) about the magnetic line.
This mechanism works only for Brownian dust clouds, because it comes
into being due to the elastic gas-grain collisions to which only
thermal granules are sensitive. To be more exact, it is assumed here that the 
precession period is much shorter than a typical time during which the
grain's angular momentum alters considerably.

The suprathermally-rotating dust particles ignore the random torques 
caused by the elastic gas-grain collisions, because the timescales for
the random torques to alter the spin state are several orders of magnitude
larger than the average time between subsequent crossovers (Purcell 
1979). Still, the
dust granules do become susceptible to the random torques during the 
brief cross-overs when the granule becomes, for a short time, thermal
(i.e., slow spinning). This is the essence of the first Lazarian
mechanism of alignment, introduced in Lazarian (1995 d) under the name
of ``Cross-Over Mechanism''. Hence, the first Lazarian mechanism is
the Gold alignment generalised for suprathermal grains; the
generalisation being possible because even suprathermal granules 
become thermal for small time intervals.

The second Lazarian mechanism, termed by Lazarian (1995 d, e)
``Cross-Section Mechanism'', and studied in Lazarian \& Efroimsky
(1996) and Lazarian, Efroimsky \& Ozik (1996), is not a generalisation
of any previously known effect, but is a totally independent, very 
subtle phenomenon. Its essence can well be grasped on the intuitive
level: 
a precessing (about the magnetic line) interstellar granule will
``prefer'' to spend more time in a rotational mode of the minimal 
effective cross section. In other words, the particle has to ``find''
the preferable mean value of its precession-cone's half-angle, value that will
minimise the mean cross section. Here, the ``mean cross section'' is 
the averaged (over rotation, and then over precession) cross section of a
granule as seen by an observer looking along the direction of 
interstellar drift. It is crucial that, though the alignment is 
due to gas-grain collisions, it establishes itself not relative to the wind direction 
but relative to the magnetic line about which the spinning grain is precessing.

Now, the goal is to understand how effective this mechanism is for
the dust particles of various geometric shapes. Articles (Lazarian and 
Efroimsky 1996) and (Lazarian, Efroimsky \& Ozik 1996) addressed the 
cross-section alignment of oblate and prolate symmetrical grains,
correspondingly. In the current paper we intend to extend the
study to ellipsoidal granules of arbitrary ratios between the semiaxes.

\section{Starlight Extinction on Interstellar Dust}

A starlight beam passing through a dust nebula gets attenuated. This
process is called extinction, and it comprises two separate
phenomena: scattering and absorption. The final result of
extinction is a cooperative effect of all dust particles the ray
bypasses. 

Concerning the absorption, one may safely take for granted that it is 
taking place on different granules independently from one another. 
For scattering, though, such independence is not, generally, guaranteed. Still,
in our study we shall deal with the independent scattering solely, in that we 
shall omit the phase relations between the waves scattered by neighbouring 
grains. This is justified by the starlight not being monochromatic:
the lack of coherence in it excludes 
whatever phase-related phenomena\footnote{It may be good to measure the 
starlight polarisation at separate wavelengths, but such a project
may require a new, more sensitive, generation of observational means.}. Thence the
intensities of waves scattered from the various granules must be added without 
regard to phase. Finally, we shall be blithe about the multiple scattering,
because it brings almost nothing in the observation. Indeed, the granules are 
separated by distances exceeding their size by many orders of magnitude, and 
the optical depth of most interstellar\footnote{This is not necessarily true 
for circumstellar environments where the dust is more abundant, and radiation 
transfer is taking place.} dust clouds is well below unity.

To conclude, we shall assume the starlight extinction by different
grains to be independent and devoid of whatever phase correlations:
the appropriate losses in intensity simply add (and never return back
through radiation transfer). This reasonably simplified approach
leaves room for a 
thorough treatment of starlight attenuation by rarified dust.

\section{The Scattering, Absorption, and Extinction Cross Sections} 

The starlight-scattering cross section on a dust grain is introduced
in a pretty standard manner. Let us begin, for simplicity, with a 
monochromatic ray, and then generalise the consideration to the
natural light. If the incident radiation is a monochromatic plane wave 
of frequency $\omega$ and intensity $I_0$, the observer located at point 
$(r,\,\theta,\,\phi)$, relative to the scatterer, will register intensity 
proportional to $I_0$ and inversly proportional to $r^{-2}$:
\be
I(\theta, \phi, \omega)\;=\;D(\theta, \phi, \omega)\;\frac{I_0}{r^2}\;\;,\;
\label{3.1}
\ee
the wavelength- and angle-dependent factor $D$ having the dimension of squared 
length. From the physical viewpoint, it can have such a dimension if and only 
if it is proportional to the squared wavelength $\lambda$ or, the same, 
inversly proportional to the squared wave 
vector $k\,=\,\omega/c\;=\;2\pi/\lambda\;$:
\be
I(\theta, \phi, \omega)\;=\;F(\theta, \phi)\;\frac{I_0}{k^2\;r^2}\;
\label{3.2}
\ee
Now let us naively suppose that all
the photons entering the scattering region through area $dC_{scat}$
get scattered into the solid angle $d\Omega$ about the direction
$(\theta, \phi)$. If this, purely corpuscular,
interpretation were correct, the energy conservation law would read:
\be
I(\theta, \phi, \omega)\;d\Omega\;r^2\;=\;I_0\;dC_{scat}\;\;.\;\;
\label{3.3}
\ee
By comparing the latter with the former, we  would then arrive to the 
expression for the scattering differential cross section:
\be
\frac{dC_{scat}}{d\Omega}\;=\; \frac{F(\theta, \phi)}{k^2}\;\;\;.
\label{3.4}
\ee
The full cross section would then be given by the integral:
\be
C_{scat}(\omega)\;=\; \int\;\frac{F(\theta, \phi)}{k^2}\;d\Omega\;\;.
\label{3.5}
\ee
Needless to say, it would be physically incomplete
to interpret  $C_{scat}$ simply as the incident wavefront
area wherefrom the photons get scattered off their initial 
direction. As well known since the times of Newton and Huygens, the
corpuscular interpretation neglects the interference of the scattered 
and incident components of light. Therefore, it will, for example, fail
to correctly describe the forward-scattering. The problem is somewhat
subtle. On the one hand, the above mathematical expression for  
differential cross section formally remains correct for whatever finite value 
of the scattering angle $\;\theta\;$. Indeed, for an 
arbitrarily small, though finite angle, the observer 
potentially {\bf{can}} distinguish between the primary and the scattered
images. To that end, he will have to employ a sufficiently powerful 
telescope located at a sufficiently remote distance from the
scatterer. On the other
hand, though, the needed resolving power of the telescope must be 
achieved by increasing the size of the object lens. The finite radius of 
lenses thereby imposes a restriction on the scattering
angle values for which (\ref{3.5}) is meaningful\footnote{Suppose, our telescope is aimed at a distant star, the 
scatterer being slightly off the line of sight. In order for the
secondary image to get into the object lens, the scattered photons must be
deflected at angles not exceeding $\;R/r\;$, with $\,R\,$ and $\,r\,$ 
being the radius of the lense and the distance to the scatterer. At
the same time, the angular resolution of the lense is less than $\,
\lambda/R\,$. This results in the trivial inequality $\;\lambda / R\,<
\,R/r\;$, whence $\;R\,>\,\sqrt{\lambda\,r}\;.$}. 
Hence, the expression for full cross section $\;C_{scat}\;$ is, in
fact, of no physical interest. It corresponds to no physical
measurement, because it is pointless to carry out
the integration over too close a vicinity of $\;\theta\,=\,0\;$. Our is 
the case: with the distances grossly exceeding the device size, 
the observations of starlight are performed at effectively zero scattering
angles; so the telescope does not distinguish the forward-scattered
light from the primary wave. To account for their interference, one
has to make a simple estimate. Let the incident radiation be described by wave
\be
u_0\;=\;\sqrt{I_0(\omega)}\;exp\{\,-\,i\,k\,z\; +\; i\,\omega \,t  \}\;\;\;.
\label{3.6}
\ee
Then the scattered wave in the distant zone will read:
\be
u=\;S(\theta, \phi)\;\sqrt{I_0(\omega)}\;\frac{exp\{\,-\,i\,k\,r\; +\; i\,\omega\,t
\}}{ikr}\; =\;S(\theta, \phi)\;\frac{exp\{\,-\,i\,k\,r\; +\; i\,k \,z
\}}{ikr}\;u_0\;\;,\;
\label{3.7}
\ee
the relative amplidude $S(\theta, \phi)$ being, generally, complex. (Evidently,
$\;|S(\theta, \phi)|^2\,=\,F(\theta, \phi)\;$.) Consider the case of
forward-scattering, when $\theta = 0$. Let the telescope with
object lens' area $A$ be located far afield and register a 
combined image of the incident and forward-scattered beams. The
aperture plane being $z=const$, a point of the object lens is located at
distance
\be
r=z+\frac{x^2+y^2}{2z}
\label{3.8}
\ee
from the scatterer. The amplitude at $\theta = 0$ will be denoted by $S(0)$.
The superimposed amplitudes will, together, give:
\be
u_0 + u = u_0\;\left[1+\frac{S(0)}{ikz}
\;\exp\left(-ik\frac{\left(x^2+y^2 \right)}{2z}\right) \right] \;\;,\;\;
\label{3.9}
\ee
the second term being a small correction to the first. Hence, the
intensity incident on this point of the lens should be:
\be
|u_0 + u|^2\; \approx\; I_0(\omega)\;\left[1\; + \;2\;Re \left\{   \frac{S(0)}{ikz}
\;\exp\left(-ik\frac{\left(x^2+y^2 \right)}{2z}\right)   \right\}\right]
\label{3.10}
\ee
integration whereof over the objective will yield the total intensity
of the combined image. As well known,
\be
\int_{-\infty}^{\infty}\;\exp\left(-ikx^2/2z
\right)\;dx\;=\;\sqrt{\frac{2\pi z}{ik}}
\label{3.11}
\ee
wherefrom
\ba
\int \int |u_0 + u|^2\;
dx\,dy\;\approx\; I_0(\omega)\;\left[ A\,+\, \frac{2}{kz}\;Re\left\{\frac{S(0)}{i}
\;\frac{2\pi z}{ik}\right\}\right]=\;I_0\;\left[ A\;-\;\frac{4\pi}{k^2}\;
Re\{S(0)\}\right]
\label{3.12}
\ea
the integration being carried out over the object-lens area $A$. Since in the
second term we had negative-power exponentials, there was nothing
wrong in approximating its integral over the finite aperture by
extending its limits from $-\infty$ through
$\infty\;$: for the illustrative estimate, the small difference
between the Fresnel and Gaussian integrals is irrelevant.
 
Evidently, the second, negative term in the latter expression gives
the amount by which the light energy entering the telescope is reduced by
the scatterer. This reduction is called {\it forward-scattering 
cross-section}:
\be
C_{f-s}(\omega)\;=\;\frac{4\,\pi}{k^2}\;Re\,\{S(0)\}
\label{3.13}
\ee
Roughly speaking, the observer will get an impression that a certain 
share, $\;C_{f-s}/A\;$ of the object lens area $A$ is covered up. 
This shows the fundamental difference between the regular scattering cross
section and forward-scattering cross section: while the former is
associated with the area of the incident wavefront, the latter is
associated with the area of the observer's aperture. This profound
difference stems from the fact that the front-scattering light cannot
be physically separated from the incident wave. As agreed above, we 
consider only optically-thin clouds. This means that we totally
ignore $\;C_{scat}\;$, but do take $\;C_{f-s}\,$ into account.

Physically, it is quite obvious that absorption will come into play 
through adding some $\,C_{abs}\;$ to $\;C_{f-s}\;$. Even less light will reach
the lense, and the observed intensity will be:
\be
I_0(\omega)\;\left[A \;-\; C_{f-s}(\omega) \;- \;C_{abs}(\omega)\right]
\label{3.14}
\ee
In the end of the preceding section we agreed that the light
extinction by different granules is independent and is free from phase 
correlations: the intensity losses simply add. This will result in the
extinction cross sections of the single granules added to give
the extinction cross section of the entire cloud (for a detailed
explanation see {\it{van de Hulst}} 1957, p. 31 - 32). Finally, for
whatever real observation, the above expression must be multiplied by
the window function $\;W(\omega)\;$ of the device, and integrated over 
$\;\omega\,$. All in all, the resulting attenuation will be expressed by the
extinction cross-section:
\be
C_{ext} = \frac{\int_{0}^{\infty} 
\;W(\omega)\;I_0(\omega)\;\;{\sum_i} \left[ C_{f-s}^{\{i\}}(\omega) + C_{abs}^{\{i\}}(\omega)
\right]\;d\omega}{\int_{0}^{\infty} \;W(\omega)\;I_0(\omega)\;d\omega}\;\;,
\label{3.15}
\ee
$\it i\;$ standing for the number of a particle, and $\;C^{\{ \it i
\}}_{abs,\;f-s}\;$ for its cross-sections. If $\;W(\omega)\;$ carves out a band wherein $\;C_{f-s}^{\{i\}} +
C_{abs}^{\{i\}}\;$ depends on $\,\omega\,$ weakly, we are left with:
\be
C_{ext} \;\approx\; 
{\sum_i}\;\left( \;C_{f-s}^{\{i\}} \;+ \;C_{abs}^{\{i\}}\;\right)\;.
\label{3.16}
\ee
All the above is valid for both polarisations
independently. So the scattering, forward-scattering, absorption, and
extinction cross sections may be introduced for the two polarisations
separately.

\section{The Measure of Alignment, and Its Relation to Observable Quantities.} 

In the Introduction we explained what it means for an interstellar
grain to be aligned. For the effect to be quantified, it should be endowed with
some reasonable measure, one that would interconnect the dust
dynamics with the starlight-polarisation degree.

Linear polarisation essentially means that, if the ray propagates
in the $\,z_o\,$ direction, there exist two (orthogonal to $\,z_o\,$ and to one
another) directions, $\;x_o\;$ and $\;y_o\;$, appropriate to the maximal 
and minimal magnitudes of the electric field in this ray. The
subscript ``$o$'' signifies the observer's frame. The
question now is: how will these maximal and minimal magnitudes 
$\;E_{x}^{o}\;$ and $\;E_{y}^{o}\;$ (or, equivalently, the maximal 
and minimal intensities, $\;{E_{x}^{o}}^2\;$ and $\;{E_{y}^{o}}^2\;$)
evolve along the line of sight, within the cloud? Properly speaking, 
one should talk about the ensemble-averages of these intensities:
$\;\langle {E_{x}^{o}}^2\rangle\;$ and $\;\langle{E_{y}^{o}}^2
\rangle\;$, the averaging being implied first over the grain orientation
(relative to its angular momentum), then over the angular-momentum's 
precession about the magnitic field and, finally, over the half-angle 
$\;\beta\;$ of the precession cone. (The direction of magnetic field
will be assumed to be constant over the line of sight, within the cloud.) 
While each of the first two averagings will be merely an integration
over a full circle, the latter averaging will involve some
distribution function for $\;\beta\;$. This distribution function
should be supplied by the detailed theory of a particular
orientational mechanism dominating the alignment process.

In neglect of the secondary scattering, the decrease in intensity, $\;dI\;$,
is proportional to the length $\;dz\;$ and to the dust-particle
density $\;n\;$ in the cloud: $\;dI\;=\;-\;C_{ext}\;n\;dz\;$, with 
$\;C_{ext}\;$ being the afore mentioned extinction cross section
(\ref{3.15}). What we observe is the intensities at the exit from the
nebula. Call them $\;\langle {E_{x}^{o}}^2\rangle\;$ and $\;\langle 
{E_{y}^{o}}^2\rangle\;$. Then
\be
\langle {E_{x}^{o}}^2\rangle\;\sim\;\exp{\left(\,-\,C^o_x\;n\;{\it
l}\,\right)}\;\;,\;\;\;\;\;
\langle {E_{y}^{o}}^2\rangle\;\sim\;\exp{\left(\,-\,C^o_y\;n\;{\it l}
\,\right)}\;\;,
\label{4.1}
\ee
$\it l\;$ being the depth of the cloud as seen by the observer, and 
$\;C^o_{x,\,y}\;$ being the mean extinction cross-sections for the two linear 
polarisations orthogonal to the observer's line of sight. 

Suppose that, prior to entering the nebula, the starlight was
unpolarised. One can
characterise the polarising ability of the cloud by the measured flux
intensity $\;I\;$ as a function of the angle of rotation of some
analysing element of the telescope. In practice, they rather employ a
relative measure, $\;P_{ext}\;$, which is the degree of polarisation due to
selective extinction (Hildebrand 1988):
\be
P_{ext}\;\equiv\;\frac{I_x\,-\,I_y}{I_x\,+\,I_y}\;=\;\frac{
\langle {E_{x}^{o}}^2 \rangle \,-\, \langle {E_{y}^{o}}^2 \rangle 
}{
\langle {E_{x}^{o}}^2 \rangle \,+\, \langle {E_{y}^{o}}^2 \rangle
}\;
\label{4.2}
\ee
As follows from the above formulae,
\ba
\nonumber
P_{ext}\;=\;
\frac{
\exp \left(\,-\,C^o_x\;n\;{\it l}\,\right) \,-\,
\exp \left(\,-\,C^o_y\;n\;{\it l}\,\right)
}{
\exp \left(\,-\,C^o_x\;n\;{\it l}\,\right) \,+\,
\exp \left(\,-\,C^o_y\;n\;{\it l}\,\right) }\;\approx
\ea
\ba
\tanh\,\left[\; \frac{C^o_x\;n\;{\it l}\;-\;C^o_y\;n\;{\it l}}{2}\;\right]\;\;\approx\;
\;\frac{C^o_x\;n\;{\it l}\;-\;C^o_y\;n\;{\it l}}{2}\;\;,
\label{4.3}
\ea
the approximation being valid for $\;P_{ext}\,\ll\,1 \;$ (with no need
to assume that
$\;C^o\;n\;{\it l}\;\ll\;1\;$). One can also introduce the 
polarisation per optical depth:
\be
\frac{P_{ext}}{\left(C^o_x\;n\;{\it l}\;+\;C^o_y\;n\;{\it
l}\right)/2}\;=\;\frac{C^o_x\;-\;C^o_y}{C^o_x\;+\;C^o_y}
\label{4.4}
\ee
No matter what measure of polarisation one prefers, this measure
involves the difference $\;(C^o_x\;-\;C^o_y)\;$ that depends on the
extinction properties of a single grain and on the degree of alignment 
in the cloud. The topic was addressed by many. A brief conclusion that 
saves type will be: no matter which
alignment mechanism dominates, the difference $\;(C^o_x\;-\;C^o_y)\;$  
should be expressed as a function of a single granule's extinction 
cross sections and of the magnetic field direction (relative to the
line of sight). Naturally, the said difference will 
be also a functional of the precession-cone half-angle disribution. 
This half-angle, often denoted as $\;\beta\;$, comprises the 
angular separation between the magnetic field and the particle's angular 
momentum precessing thereabout. The statistical distribution of 
$\;\beta\;$  over the ensemble depends upon the dominating
orientational mechanism(s), and its calculation is a technical issue 
which sometimes is extremely laborious and involves numerics.  

The expressions for $\;C^o_x\;$ and $\;C^o_y\;$ in terms of the afore
mentioned arguments were given, for the cases of oblate and prolate
symmetrical granules, in  Greenberg 1968, Purcell \& Spitzer 1971, Lee 
\& Draine 1985, Hildebrand 1988, and Roberge \& Lazarian 1999. To
fulfil the goal of our study, we must generalise those results for 
the case of triaxial ellipsoid. To this end, we introduce the
extinction cross sections $\;C_X\;$, $\;C_Y\;$, $\;C_Z\;$ of the
grain, for light polarised along its minimal ($X$), middle 
($Y$) and maximal ($Z$) inertia axes\footnote{These cross
sections characterise particular species of dust. Computation of
the granule extinction cross sections is comprehensively discussed by van de
Hulst 1957. (See also Martin (1974) and Draine \& Lee (1984).)}. 
The calculation, presented in Appendix A, results in the
following relation:
\be
C^o_x\;-\;C^o_y\;=\;\left[\,\frac{C_X\,+\,C_Y}{2}\;-\;C_Z
\,\right]\;R\;\cos^2 \gamma\;\;\;,
\label{4.5}
\ee
where $\;\gamma\;$ is the angle between the magnetic field and the plane of
sky (Fig. 1), and $\;R\;$ is the so-called Rayleigh reduction factor
defined as
\be
R\;\equiv\;\frac{3}{2}\;\left({\Big{\langle}} \cos^2 \beta
{\Big{\rangle}}\;-\;\frac{1}{3}\right)\;\;,
\label{4.6}
\ee
$\beta\;$ being the half-angle of the precession cone described by the grain's
angular momentum about the magnetic line (Fig.2). As already mentioned, this 
angle is not the same for all  
\break
\begin{figure}
\centerline{\epsfxsize=3.5in\epsfbox{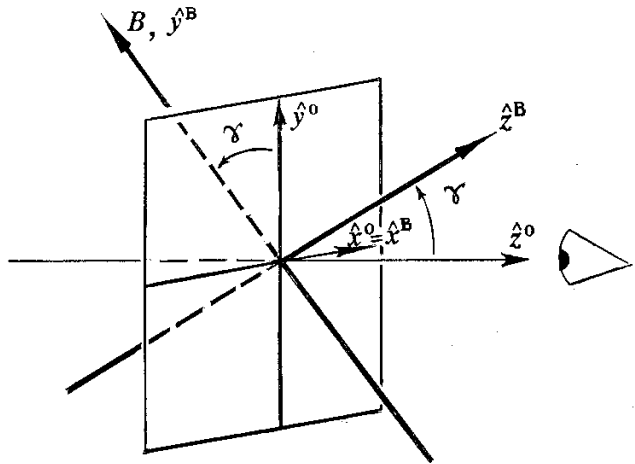}}
\bigskip
\caption{The line of sight, the plane of sky, and the direction of
magnetic field. Axes $\;y^o\;$ and $\;z^{\tiny B}\;$ are chosen to
belong to the plane defined by the line of sight and magnetic line.\\
~\\}
\end{figure}
\begin{figure}
\centerline{\epsfxsize=3.5in\epsfbox{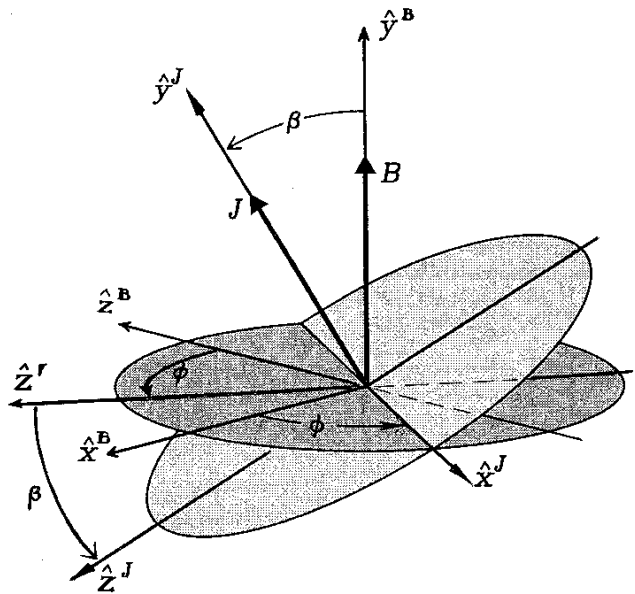}}
\bigskip
\caption{Relative positions of the coordinate systems associated with
the magnetic line and with the angular momentum. The latter system is
obtained from the former through a rotation about axis $\;y^{\tiny
B}\;$, by angle $\;\phi\;$, and a subsequent rotation about axis
$\;x^{\tiny J}\;$, by angle $\;\beta\;$.}
\end{figure}
\pagebreak
\noindent
grains, but obeys some statistical distribution determined by a 
particular physical setting. The above formula for $\;R\;$ was obtained in 
assumption of principal rotation (when the alignment of the principal axis with
the angular momentum is promptly enforced), valid for suprathermally rotating
grains (Lazarian \& Efroimsky 1999). In the thermal case, when angle 
$\;\theta\;$ between the angular momentum and the major-inertia axis is 
nonzero, the Rayleigh factor would look more complicated:
\be
R\;=\;{\Big{\langle}}  \;\;
\frac{3}{2}\;\left( \cos^2 \beta\;-\;\frac{1}{3} \right) \;\;\;
\frac{3}{2}\;\left( \cos^2 \theta\;-\;\frac{1}{3} \right) \;\;
{\Big{\rangle}} \;\;\;\;\;.
\label{4.7}
\ee
Returning to (\ref{4.6}), we would point out the evident
fact that, in the absence of alignment, the ensemble average of
$\;\cos^2 \beta\;$ is equivelent to averaging simply over the solid angle.
It entails: $\;\langle \cos^2 \beta \rangle\;=\;1/3\;$. This nullifies
the R factor and makes the radiation unpolarised: $\;C_x^o\;=\;C_y^o\;$.

\section{Calculation of the Rayleigh reduction factor in the case
when the Cross-Section Mechanism is dominant}

The   basic idea of the cross-section orientational process (pioneered by 
Lazarian 1995d, 1995e) comes from the fact that the average time between two 
sequent crossovers is proportional to a typical lifetime of an active site 
(``Purcell rocket''). Each such site is eventually ``poisoned'' through the 
evergoing accretion of atoms brought by the interstellar wind. Emergence of new
active sites leads to cross-overs. Henceforth, the higher the accretion rate, 
the shorter the average lifetime of a typical Purcell rocket. Now, since the 
atoms adsorbed  by the surface get delivered through the gas bombardment, one
may state that the said lifetime is proportional to the rate of gas-grain 
collisions.  The latter rate, in its turn, is proportional to the gas-grain 
effective cross-section, i.e., to the averaged (over the period of precession 
about the magnetic line) cross section of a granule, as seen by an observer 
looking along the line of gas flow. To summarise: the average time between two 
sequent crossovers is proportional to an active-site's lifetime, the latter 
being proportional to the rate of gas-grain collisions, which in its turn is
proportional to the effective cross section of the precessing granule in the 
flow. All in all, the dust particle will spend longer times in rotational 
states of smaller effective cross section. The cross-section mechanism
is essential for the both cases of rapid and slow flows\footnote{For
high ($\;>\,2\;\,km/s\,$) relative velocities, the cross-section mechanism
coerces the grain to align in the same direction as the cross-over
mechanism does (Lazarian 1995d). For slower flows, the stochastic torques
produced by the Purcell rockets exceed the torques caused by
collisions with gas. Thence, no considerable alignemnt will arise during the 
flip-overs. This makes the role of the cross-over mechanism marginal. 
Therefore, one may expect that the cross-section mechanism will
dominate in slow flows. Still, the flow should be, at least, mildly
supersonic, in order for the drift to dominate over the stochastic
motion of individual atoms}.

This line of reasoning, developed by Lazarian, brings into play
several time scales. One is the 
period $\;t_{rot}\;$ of grain spin about its own rotation axis. The other 
is the wobble period $\;t_{wobble}\;$, i.e., the period of precession of the 
angular velocity $\;\bf \Omega\;$ and major-inertia axis $\;Z\;$ about the 
angular momentum $\;\bf J\;$. Third is the period $\,T\,$ of precession of 
$\;\bf J\;$ about the magnetic field $\; \bf B\;$. The fourth one is the mean 
time $\;\tau\;$ between subsequent flip-overs of the granule. 
                                  
Suprathermal rotation is swift: time $\;t_{rot}\;$ is much shorter than the 
other time scales involved. Time $\;t_{wobble}\;$ is irrelevant in the 
suprathermal case, because in this case we neglect the wobble: as explained
in Section I, the grain's magnetic moment, the angular-velocity vector, and the
maxumum-inertia axis are all aligned along the angular momentum, and they 
all precess about $\;\bf B\;$, always remaining parallel to one another. 
The rate of this precession about $\;\bf B\;$ is slower than the granule's
rotation, but still rapid enough: as mentioned in Section I, the precession 
period $\,T\,$ is much shorter than a typical interval $\;\tau\;$ between 
cross-overs. A cross-over of a spinning particle can happen for one (or 
both) of the two reasons: (1) spin damping through collisions with gas atoms, 
and (2) grain resurfacing that alters positions of active cites. Without 
going into detailed dynamics, let us assume that on the average a cross-over 
takes  place after the granule experiences $\;N\;$ collisions. Suppose, this  
amount of collisions is achieved during time $\;\tau\;$:
\be
N\;=\;\int_{0}^{\tau}\;n\;u\;{\big{\langle}} S {\big{\rangle}}_{\eta}\;dt
\label{5.1}
\ee
$n\,$ being the density of atoms, $\,u\,$ being the speed of gas flow, and 
$\,{\big{\langle}} S {\big{\rangle}}_{\eta}\,$ being the cross-section of
the gas-grain interaction, averaged over the grain's (principal) rotation about
$\,\bf J\;$: 
\be
{\big{\langle}} S {\big{\rangle}}_{\eta}\;=\;\frac{1}{2\pi}\;
\int^{2\pi}_{0}\;S(\Phi,\;\beta, \phi,\;\eta)\;d \eta\;\;,
\label{5.2}
\ee 
the angles $\;\Phi\;$, $\;\beta\;$, $\;\phi\;$, and $\;\eta\;$ being
as on Figures 2 - 3.
Even after this averaging, $\;{\big{\langle}} S {\big{\rangle}}\;$ remains 
time-dependent, because of the precession of $\;\bf J\;$ about $\;\bf B\;$. For
times $\;\tau\;$ much longer than the precession period $\;T\;$,
\be
N\;=\;\frac{\tau}{T}\;\int_{0}^{T}\;n\;u\;{\big{\langle}} S {\big{\rangle}}_{\eta}\;dt\;\;,
\label{5.3}
\ee
whence
\be
\tau\;=\;
\frac{N}{    n\,u\,(1/T)\,   
{\int_{0}^{T}} 
\, {\big{\langle}} S
{\big{\rangle}}_{\eta}\;dt}
\label{5.4}
\ee
It can be also re-written as
\be
\tau\;=\;\frac{N}{    n\,u\,(1/2 \pi )\,\int_{0}^{2\pi} \, {\big{\langle}} S
{\big{\rangle}}_{\eta}\;d \phi }\;=\;\frac{N}{n\;u\;(1/2\pi)\;\langle 
\;\langle S
\rangle_{\eta}\rangle_{\phi}}\;\;,
\label{5.5}
\ee
$\,\phi\;$ being the angle as on Fig.2: during the precession period it changes
from zero to $\;2\pi \;$. The usefulness of the latter expression becomes 
evident if one recalls that the probability to find a granule in a
certain spin state is proportional to the time it stays there. After 
averaging over the rotation about $\;\bf J\;$, and after a further averaging
over the precession, the so
averaged spin state depends upon two agruments: the
precession-cone half-angle $\;\beta\;$, and the angle $\;\Phi\;$
between the magnetic line and gas drift
(see Fig.3). Thence, what the above formula gives us is the (not yet normalised) distribution  
of the (doubly averaged) spin states over $\;\beta\;$ (angle
$\;\Phi\;$ being fixed and playing the role of parameter). What then
remains is simply to normalise, i.e., to divide $\;\tau\;$
by its integrand over the solid angle. 
Thus we come to distribution
\be
p_{\mbox{\tiny }}(\beta)\;=\;\frac{1}{C}\;\frac{1}{\langle 
\;\langle S
\rangle_{\eta}\rangle_{\phi}}
\label{5.6}
\ee
with the normalisation constant equal to
\be
C\;=\;\int_{0}^{2\pi}\;d\beta\;\sin \beta\;\frac{1}{\langle 
\;\langle S
\rangle_{\eta}\rangle_{\phi}} \;\;
\label{5.7}
\ee
and the average defined as
\be
\langle 
\;\langle S
\rangle_{\eta}\rangle_{\phi}\;=\;\int_{0}^{2\pi}d\phi\;\int_{0}^{2\pi}d\eta\;\;S
\label{5.8}
\ee
This distribution being on our hands, the Rayleigh reduction factor is 
straightforward:
\be
R\;=\;\;\frac{3}{2}\;\left({\Big{\langle}} \cos^2 \beta
{\Big{\rangle}}\;-\;\frac{1}{3}\right)\;=\;\frac{3}{2}\;\left(
\int_{0}^{2\pi}\;d\beta\;\sin \beta \;\, p_{\mbox{\tiny
}}(\beta)\;\,\cos^2 \beta\;-\;\frac{1}{3} \right)\;
\label{5.9}
\ee
This is where the physics ends and mathematics begins.

\section{The distribution of $\;\beta\;$ for triaxial-ellipsoid-shaped granules}

In this section we shall compute the distribution 
$\;p_{\mbox{\tiny }}(\beta)\;$ which will, in fact, be a function of
several arguments. One is, naturally, $\;\beta\;$ itself. Another
will be the angle $\;\Psi\;$ between the magnetic line and the gas flow. 
Above that, $\;p_{\mbox{\tiny }}\;$ will depend upon the geometry of grain
(in assumption that all particles in the cloud are alike).
       
To make the section readable, we shall move most part of the calculations to 
the Appendix. Remarkably, these calculations include not just exercises in 
geometry but also elements of the variational calculus, a heavy-duty tool 
seldom required in astrophysics.
       
We shall model the cosmic-dust particle with an ellipsoid of semiaxes $\;a\,
\ge\,b\,\ge\,c\;$. Three coordinate systems will be employed. Frame $\;(X,\,Y\,
Z)\;$ will be associated with semiaxes $\;a,\,b,\,c\;$, correspondingly. As we 
know from Section I, in the suprathermal case the major-inertia axis and the 
angular velocity are aligned with the angular momentum $\;\bf J\;$. So vector 
$\;\bf J\;$ is pointing along $\;Z\;$. Another frame, $\;(x_B\,,\;y_B\,,\;z_B)
\,$ will be associated with the magnetic field $\;\bf B\;$, axis $\;y_B\;$ 
pointing along $\;\bf B\;\,$ (Fig.2). The direction of gas flow will be denoted
by unit vector $\;\bf f\;$ with components $\;(X_f,\,Y_f,\,Z_f)\;$ in the body 
frame. The angle between $\;\bf f\;$ and axis $\;Z\;$ will be called $\;\alpha
\;\,$ (Fig.3). The third coordinate system needed, $\;(x_J\,,\;y_J\,,\;z_J)\,$ 
\break
\begin{figure}
\centerline{\epsfxsize=3.5in\epsfbox{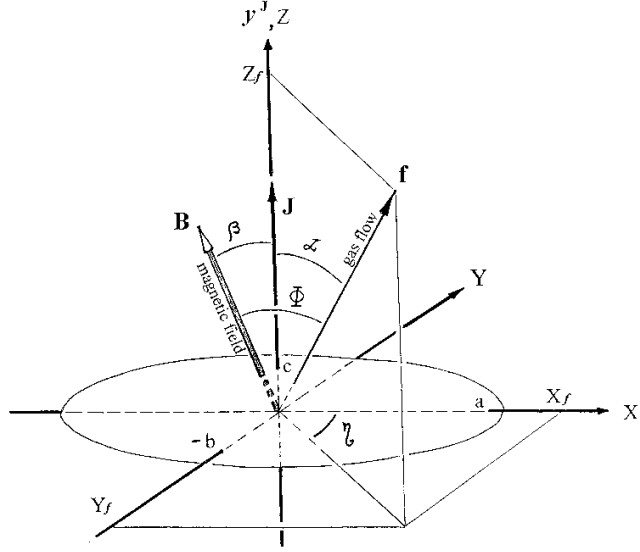}}
\bigskip
\caption{Relative orientation of the gas flow (depicted by vector $\;\bf f\;$),
the\\
$\left.\right.$ ~~~~~~~ ~~  ~~ ellipsoid's principal axes, and its
angular momentum $\;\bf 
J\;$. In the\\$\left.\right.$ ~~~~~~~ ~~  ~~ principal rotation state
the major-inertia axis $\;Z\;$ of the 
body\\$\left.\right.$ ~~~~~~~ ~~  ~~ (which is its shortest dimension) is aligned with $\;\bf J\;$.\\}
\end{figure}
\noindent
will be associated with the angular-momentum vector so that $\;y_J\;$
will be parallel to $\;\bf J\;$ and, therefore, to $\;Z\;$. Then axes $\;x_J\;$
and $\;z_J\;$ will belong to plane $\;(X,\,Y)\;$. One is free to
parametrise the rapid rotation of the granule about its major-inertia
axis  $\;Z\;$ by angle $\;\eta\;$ between the least-inertia axis
$\;X\;$ and the gas flow projection on the plane $\;(X,\,Y)\;$.
The suprathermal spin about $\;\bf J\;$ is a much 
faster process than the precession of $\;\bf J\;$ about the magnetic line. 
Therefore, while averaging over $\;\eta\;$, one may assume the orientation of 
$\;\bf J\;$ being unaltered. This means that during several rotations of the 
granule (about $\;\bf J\;$) the angle $\;\alpha\;$ between $\;\bf J\;$ and the
gas flow may be assumed unchanged.
We shall also need angle $\;\Phi\;$ between the gas flow and the magnetic 
field, and angle $\;\beta\;$ between the magnetic field and the angular 
momentum. Finally, we shall parametrise the precession of $\;\bf J\;$
about $\;\bf B\;$ by angle $\;\phi\;\,$ (Fig.2). This angle is constituted by 
axis $\;z_B\;$ and axis $\;z'\;$ (which is the projection of $\;\bf J\;$ on the
plane perpendicular to $\;\bf B\;$). Without loss of generality, one can direct
axis $\;x_B\;$ along the gas-flow projection on the plane perpendicular to 
$\;\bf B\;$.

As evident from Fig. 4, angles $\;\alpha\;$, $\;\beta\;$, 
$\;\Phi\;$ and $\;\phi\;$ are not all independent. They obey the (proven in 
Appendix B) relation:
\be
\cos \alpha\;=\;\cos \Phi\;\cos \beta\;+\;\sin \Phi\;\sin \beta\;\sin \phi\;\;.
\label{6.1}
\ee
\begin{figure}
\centerline{\epsfxsize=3.5in\epsfbox{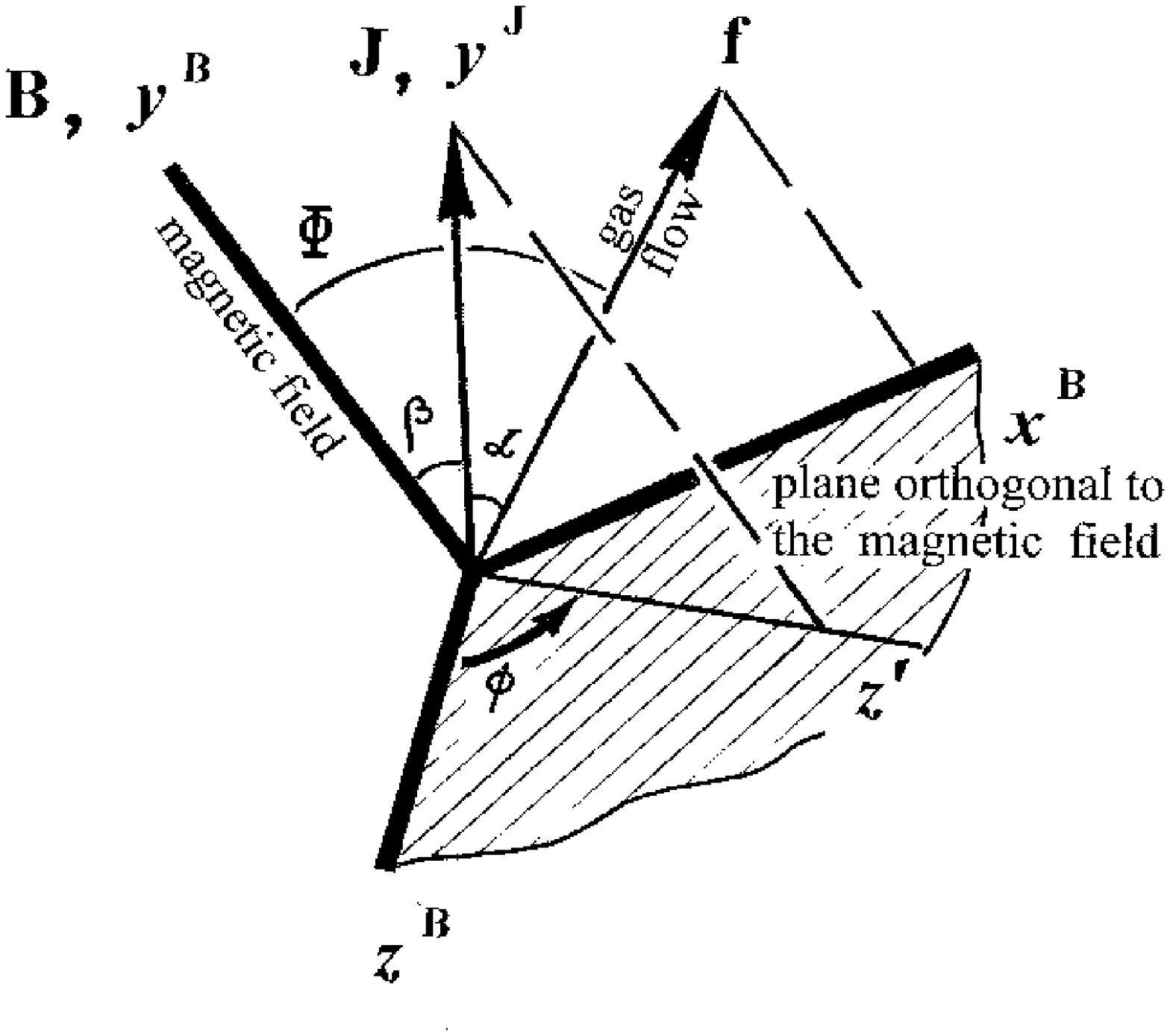}}
\bigskip
\caption{Axis $\;y^{\tiny B}\;$ is chosen to point along the magnetic
line. Axis $\;x^{\tiny B}\;$ is chosen to\\
$\left.\right.$ ~~~~~~~ ~~ ~~ point along the projection
of the flow on the plane orthogonal to magnetic\\ $\left.\right.$ ~~~~~~~ ~~ ~~ field $\;\bf B\;$.
Auxiliary axis $\;z'\;$ points along the projection of the angular
mo-\\$\left.\right.$ ~~~~~~~ ~~ ~~ 
mentum $\;\bf J\;$ on the said plane.}
\end{figure}
\noindent
The gas flow speeding by the ellipsoid defines an elliptic curve
bounding area $\;\Sigma\;$ hatched on Fig. 5 and depicted by a
thick solid line on Fig. 6. Evidently, 
\be
\Sigma\;=\;\pi\;|{\robold}_{min}\;\times\;{\robold}_{max}|\;=\;\pi\;{\rho}_{min}\;{\rho}_{max}\;\;\;,
\label{6.2}
\ee
${\robold}_{min}\;$ and $\;{\robold}_{max}\;$ being the semiaxes of hatched 
ellipse. Projection of  $\;\Sigma\;$ on the plane orthogonal to the gas flow 
will give us the cross section $\;S\;$ of the granule as seen by the
observer looking at it along the line of wind. As shown in Appendix B, 
\be
S\;=\;\pi\;\,\mid n_x\;\cos \eta\;\sin \alpha \;+\;n_y\;\sin
\eta\;\sin \alpha\;+\;n_z\;\cos \alpha\,\mid\;\;\;,
\label{6.3}
\ee
where the body-frame components of the auxiliary vector $\bf n \equiv
{\robold}_{min}\times{\robold}_{max}$ are expressed by
\begin{figure}
\centerline{\epsfxsize=3.5in\epsfbox{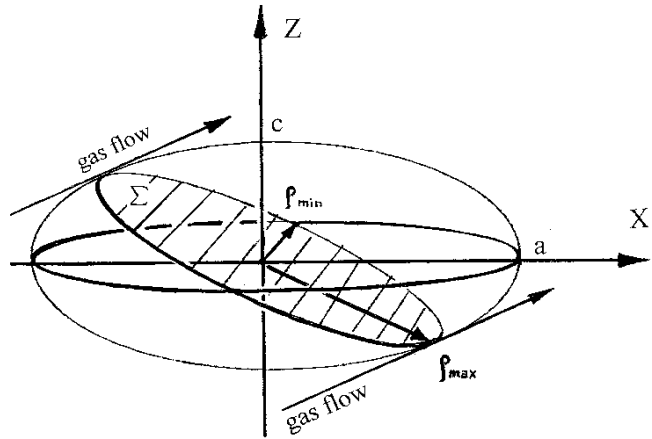}}
\bigskip
\caption{Gas flow passing by a granule is tangential to its surface in
certain points.\\$\left.\right.$ ~~~~~~~ ~~ ~~  Altogether such points constitute an ellipse whose
interior is hatched. Its\\$\left.\right.$ ~~~~~~~ ~~ ~~  semiaxes are
vectors $\;\robold_{min}\;$ and $\;\robold_{max}\;$.\\
$\left.\right.$\\}
\end{figure}
\begin{figure}
\centerline{\epsfxsize=3.5in\epsfbox{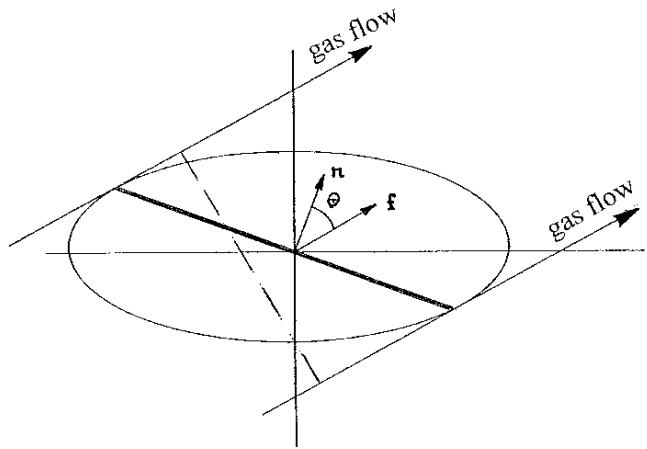}}
\bigskip
\caption{A granule in the gas flow. The thick solid line denotes cross
section $\;\Sigma\;$ (the\\$\left.\right.$ ~~~~~~~ ~~ ~~ one that is hatched on Fig. 5). ~~The dashed
line is the cross section $\;S\;$ of\\$\left.\right.$ ~~~~~~~ ~~ ~~   the granule relative to the
flow. Unit vectors normal to these cross sections, \\$\left.\right.$ ~~~~~~~ ~~ ~~ $\bf n\;$  and $\;\bf
f\;$, are separated by angle $\;\theta\;$.\\}
\end{figure}
\pagebreak
\be
n_x\;=\;(\rho_{min})_y\;(\rho_{max})_z\;-\;(\rho_{min})_z\;(\rho_{max})_y\;\;
\label{6.4}
\ee
and its cyclic transpositions. The components of $\;\robold_{min}\;$ and 
$\;\robold_{max}\;$ depend upon the lengths $\;a\;$, $\;b\;$, $\;c\;$ of the 
semiaxes, and upon their orientation relative to the gas flow, i.e.,
upon angles $\;\eta\;$ and $\;\alpha\;$. Angle $\;\alpha\;$, in its
turn, depends upon $\;\beta\;$, $\;\Phi\;$ and $\;\phi\;$. All this
will, eventually, enable us to express $\;S\;$ via 
$\;a\;$, $\;b\;$, $\;c\;$, $\;\beta\;$, $\;\Phi\;$, $\;\phi\;$ and $\;\eta\;$.
After that we shall average over $\;\eta\;$ (i.e., over the principal
rotation about $\;\bf J\;$) and over $\;\phi\;$ (i.e., over the
precession of  $\;\bf J\;$ about  $\;\bf B\;$). It will give us the
distribution (\ref{5.6}) over $\;\beta\;$, baring dependence also upon
$\;a\;$, $\;b\;$, $\;c\;$ and $\;\Phi\;$ as parameters.

Calculation of the components of vector $\;\bf n\;$ is presented in Appendix C.
Plugging of these in (\ref{6.3}) entails the following expression for the cross
section of the grain placed in the flow:
\ba
\nonumber
S\,=\,\frac{\pi}{4}\,\mu_1\,\mu_2\,\mid\sin^2 \alpha\,\cos \alpha\,\sin \eta \,\cos\eta\mid\,\left[
\frac{1}{\left(b^2\,-\,\lambda_2
\right)\,\left(c^2\,-\,\lambda_1\right)}\,-\,
\frac{1}{\left(c^2\,-\,\lambda_2 \right)\,\left(b^2\,-\,\lambda_1
\right)}\right.\;\;\;\\
\;+\;
\frac{1}{\left(c^2\,-\,\lambda_2 \right)\,\left(a^2\,-\,\lambda_1
\right)}\;-\;\frac{1}{\left(a^2\,-\,\lambda_2
\right)\,\left(c^2\,-\,\lambda_1 \right)}\;\;\;
\label{6.5}\\
\nonumber
\;\;\;\;\;\;\;\;\;\;\;\;\;\;\;\;\;\;\;\;\left.\;+\;
\frac{1}{\left(a^2\,-\,\lambda_2 \right)\,\left(b^2\,-\,\lambda_1\right)}\;-\;\frac{1}{\left(b^2\,-\,\lambda_2 \right)\,\left(a^2\,-\,\lambda_1\right)}
\;\right]
\ea
where
\ba
\mu_1\;=\;2\;\left( \left(\frac{\sin\alpha\;\cos\eta}{a\;\left(a^2\,-\,\lambda_1\right)}\right)^2\;+\;
                   \left(\frac{\sin\alpha\;\sin\eta}{b\;\left(b^2\,-\,\lambda_1\right)}\right)^2\;+\;
                   \left(\frac{\cos\alpha}{c\;\left(c^2\,-\,\lambda_1\right)}\right)^2  \right)^{-1/2}\;\;,\;\;\;\;
\label{6.6}
\ea
\ba
\mu_2\;=\;2\;\left(\left(\frac{\sin\alpha\;\cos\eta}{a\;\left(a^2\,-\,\lambda_2\right)}\right)^2\;+\;
                   \left(\frac{\sin\alpha\;\sin\eta}{b\;\left(b^2\,-\,\lambda_2\right)}\right)^2\;+\;
                   \left(\frac{\cos\alpha}{c\;\left(c^2\,-\,\lambda_2\right)}\right)^2   \right)^{-1/2}\;\;\;,
\label{6.7}
\ea
\be
\lambda_{1,2}\;=\;\frac{-\;Q\;\pm\;\sqrt{Q^2\;-\;4\;P\;R}}{2\;P}\;\;\;,
\label{6.8}
\ee
\be
P\;=\;\left(\frac{\sin\alpha\;\cos\eta}{a}\right)^2\;+\;\left(\frac{\sin\alpha\;\sin\eta}{b}\right)^2\;+\;
\left(\frac{\cos \alpha}{c}\right)^2\;\;,\;\;\;\;
\label{6.9}
\ee
\be
Q=-\left[
\left(\frac{\sin\alpha\;\cos\eta}{a}\right)^2\left(b^2\,+\,c^2\right)\,+\,
\left(\frac{\sin\alpha\;\sin\eta}{b}\right)^2\left(c^2\,+\,a^2\right)\,+\,
\left(\frac{\cos \alpha}{c}\right)^2\left(a^2\;+\;b^2\right)
\right]
\label{6.10}
\ee
\be
R\;=\;\left(\frac{\sin\alpha\;\cos\eta}{a}\right)^2\;b^2\;c^2\;+\;
           \left(\frac{\sin\alpha\;\sin\eta}{b}\right)^2\;c^2\;a^2\;+\;
           \left(\frac{\cos \alpha}{c}\right)^2\;a^2\;b^2\;\;\;.
\label{6.11}
\ee
What remains is to average $\;S\;$ over $\;\eta\;$ and $\;\phi\;$
(using formula (\ref{6.1})), and to plug the inverse of $\;\langle\;\langle S 
\rangle_{\eta}\rangle_{\phi}
\;$ in the expression $\;(\ref{5.6}) - (\ref{5.8})\;$ for the distribution
$\;p(\beta)\;$. The latter will give us, through (\ref{5.9}), the 
Rayleigh reduction factor as a function of the granule dimensions $\;a,\,b,\,c
\;$, and of the angle $\;\Phi\;$ between the magnetic field and the gas drift.

This work can be performed only numerically, and 
must be carried out with a great care. The difficulty emerges from the fact
that the denominators of the terms in square brackets in the above
expression (\ref{6.5}) for $\;S\;$ vanish at certain values of the angles and at
certain values of $\;a,\,b,\,c\;$. Fortunately, this is fully
compensated by the multipliers $\;\mu_1\,\mu_2\,|\sin^2 
\alpha\,\cos \alpha\,\sin \eta \,\cos\eta \,|\; $ (which is most natural,
for the area of a shadow cast by a smooth granule cannot have singularities).
Still, when it comes to numerics, the mentioned issue demands much attention.

\section{Results and their physical interpretation.}

The results of computation are presented on Figures 7 - 9. 
As expected, the diagrammes on all three pictures show the
symmetry that corresponds to the invariance under inversion of
the magnetic field direction. For another easy check-up, we
see that on all the diagrammes the Rayleigh reduction factor vanishes in
the limit of spherical grains.  

We see that
the cross-section Lazarian alignment is intensive for oblate granules (Fig. 7),
and approaches its maximum in the limit of ``flat flake'' shape. The alignment is maximal when the flow is paraller (or antiparallel)
to the magnetic line, or is perpendicular thereto. Between these
extremes, the alignement goes through zero. To explain this, let us
consider the simple case of flate flake, and recall that the granule,
roughly speaking, ``wants'' to minimise its 
(averaged over rotation and precession) cross-section as ``seen'' by the flow.
When the flow is parallel (or antiparallel) to the magnetic line
about which the grain precesses, the average cross section is minimal
if the flake has its precession-cone half-angle 
\clearpage
\begin{figure}
\begin{picture}(441,216)
\includegraphics{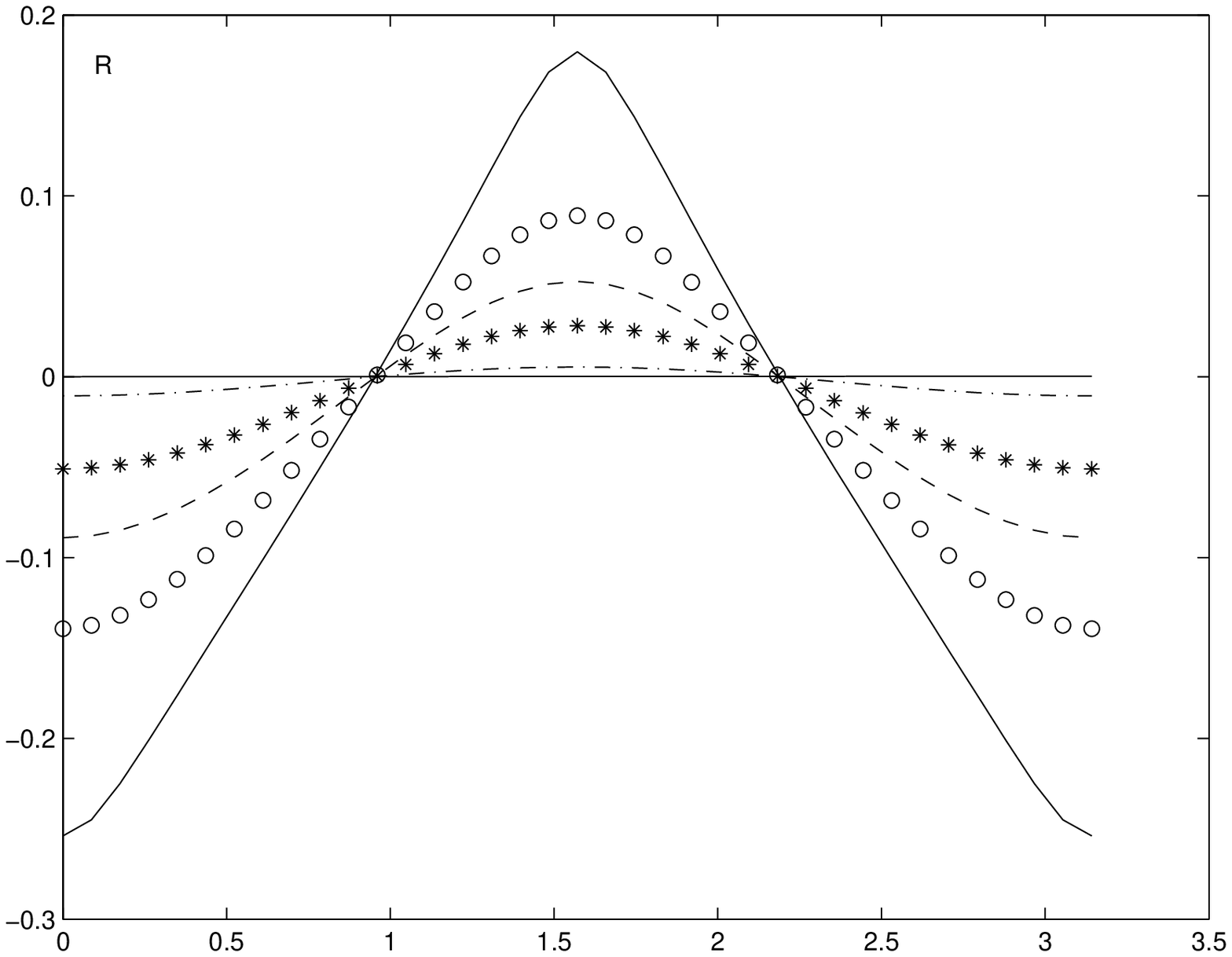}
\end{picture}
\\
\caption[]{The Rayleigh reduction factor R as a function of the angle 
$\;\Phi\;$ between\\ $\left.\right.$~~~~~~~~~~the magnetic 
field and the gas flow. The
case of oblate symmetrical granules:\\ 
$\left.\right.$~~~~~~~~~~the solid line corresponds to the semiaxes ratio
$\;1\;:\;1\;:\;1/10\;\;$ (flat ``flakes''),\\ 
$\left.\right.$~~~~~~~~~~the circle line corresponds to $\;1:1:1/3\;,\;\;\;$
the dashed line corresponds to $\;1:1:1/2\;,$\\
$\left.\right.$~~~~~~~~~~the star line corresponds to $\;1:1:2/3\;,\;\;\;$
the dash-dot line corresponds to $\;1:1:0.9\;$.\\}
\end{figure}
\noindent
$\;\beta\;$ close to
$\;\pi/2\;$ (and its square cosine close to nil). The $R$ factor will
be negative.
When the drift is
orthogonal to the magnetic field, the flake ``prefers'' to minimise
its average cross-section by rotating at $\;\beta\;$ close to zero
(with its squared cosine close to unity). The $R$ factor will be
positive\footnote{Analytical treatment is possible in the cases of $\;\Phi
\;=\;0\;$ (when radiation-pushed grains follow the magnetic line) and
of $\;\Phi\;=\;\pi/2\;$ (when the grains are subject to Alfvenic waves
or ambipolar diffusion). For details see Lazarian \& Efroimsky
(1996)}. Therefore, $R$ passes through zero when the angle takes some 
intermediate value, that may be different for different ratios of
semiaxes. Contrary to the expectations, though, this value, 
$\;\Phi_{0}^{\tiny{oblate}}\;$, bares no dependence on the semiaxes' ratio. \\
 
\begin{figure}
\begin{picture}(441,216)
\includegraphics{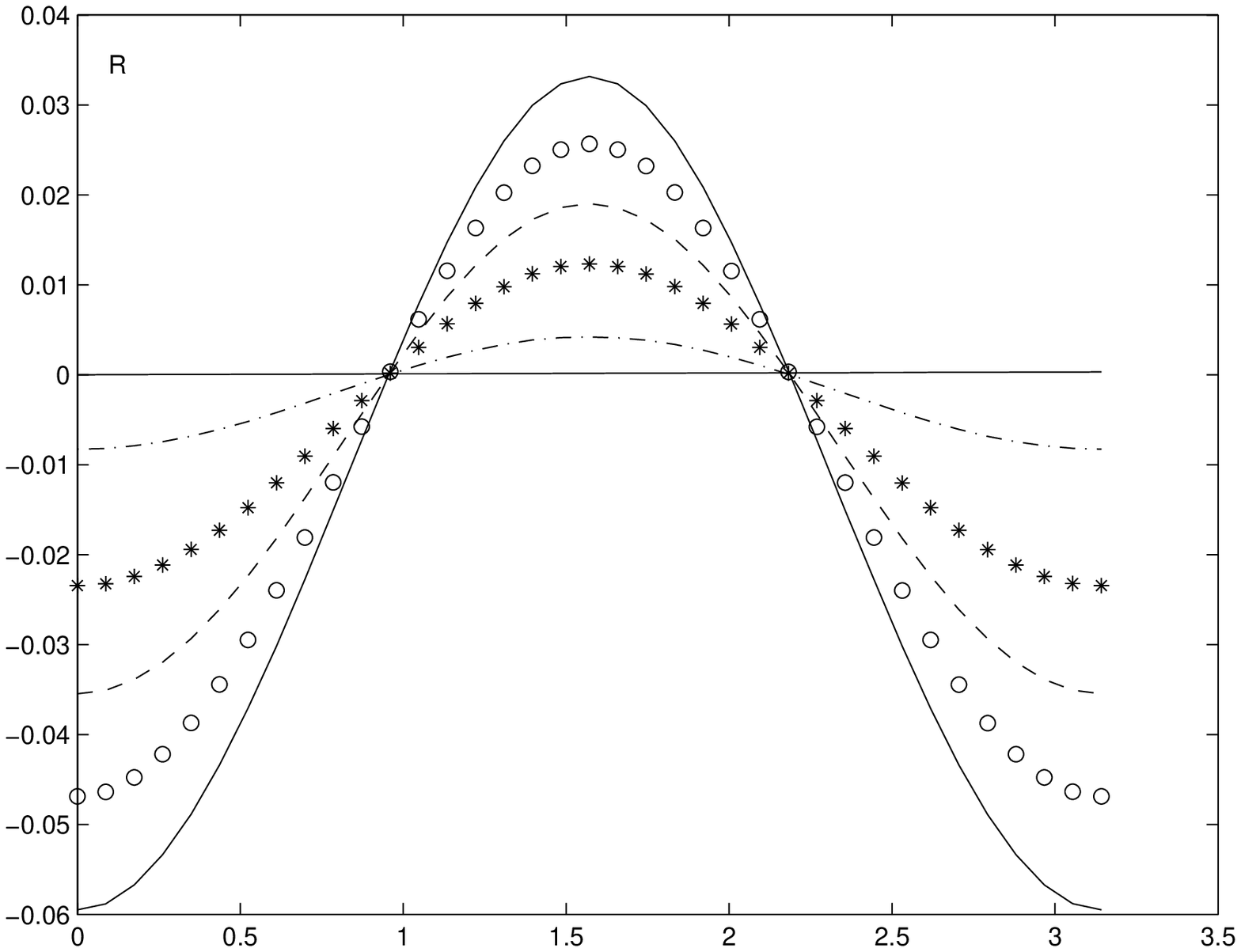}
\end{picture}
\caption[]{The Rayleigh reduction factor R as a function of the angle 
$\;\Phi\;$ between\\ $\left.\right.$~~~~~~~~~~the magnetic field and the gas flow. The
case of prolate symmetrical granules:\\ 
$\left.\right.$~~~~~~~~~~the solid line corresponds to the semiaxes
ratio $\;1\;:\;1/10\;:\;1/10\;$ (elongated ``rods''),\\ 
$\left.\right.$~~~~~~~~~~the circle line corresponds to $\;1:1/3:1/3\;,\;\;\;$ 
the dashed line corresponds to $\;1:1/2:1/2\;$,\\
$\left.\right.$~~~~~~~~~~the star line corresponds to $\;1:2/3:2/3\;,\;\;\;$
the dash-dot line corresponds to $\;1:0.9:0.9\;$.\\}
\end{figure}
As evident from Fig.8, the effect is much weaker 
in the case of prolate grains\footnote{This case was
studied in (Lazarian, Efroimsky \& Ozik 1996). Our
Fig.8 is in full agreement with Fig.3 from that paper. Mind,
though, that on Fig.3 in the said paper there is a
slip of the pen: in fact, $\;\Phi\;$ was changing not from $\;0\;$ through $\;\pi /2\;$ 
but from $\;\pi /2\;$ through $\;\pi\;$.}.
The diagrammes are 
similar to those of oblate case, and the physical interpretation
is the same as above. Just as in the case of oblate geometry, the $R$ factor
goes through zero at some value of $\;\Phi\;$, that may depend on the
semiaxes' ratio. Remarkably, in this case, too, such a dependence is
absent, and all the curves cross
the horizontal axis in the same point $\;\Phi_0^{\tiny{prolate}}\;$.

Moreover, the angles $\;\Phi_0^{\tiny{prolate}}\;$ seems to
coincide with $\;\Phi_{0}^{\tiny{oblate}}\;$ and (within the limits 
imposed by the calculation error) equals
$\;\it{arccos}\,(1/\sqrt{3})\;$. Such miraculous coincidence must reflect some 
physical circumstances that are not evident at the first glance. A 
straightforward 
\begin{figure}
\begin{picture}(441,216)
\includegraphics{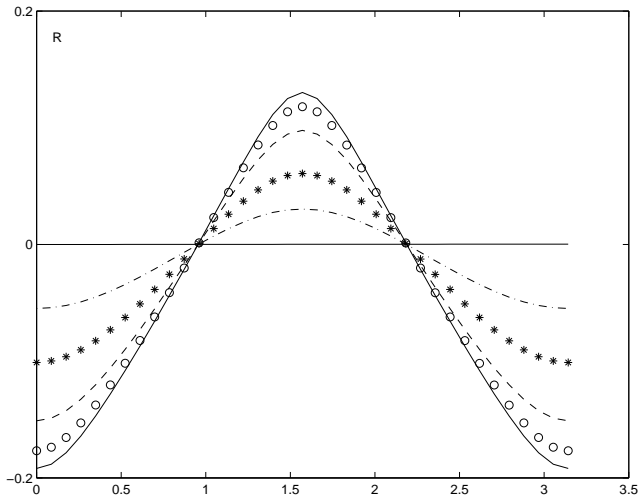}
\end{picture}
\caption[]{The Rayleigh reduction factor R as a function of the angle 
$\;\Phi\;$ between\\ $\left.\right.$~~~~~~~~~~the magnetic field and the gas flow. The
case of triaxial asymmetrical granules:\\ 
$\left.\right.$~~~~~~~~~~the solid line corresponds to the semiaxes ratio $\;1\;:\;0.9\;:\;0.2\;$;\\ 
$\left.\right.$~~~~~~~~~~the circle line corresponds to $\;1:0.7:0.2\;;\;\;\;$ 
the dashed line corresponds to $\;1:0.5:0.2\;$;\\
$\left.\right.$~~~~~~~~~~the star line corresponds to $\;1:0.3:0.2\;;\;\;\;$
the dash-dot line corresponds to $\;1:0.2:0.2\;$.\\}
\end{figure}
\noindent
analytic proof of this ``coincidence'', even in the 
simplest, oblate case, is unavailable.

The third picture, Fig. 9, accounts for the general case of a triaxial body, 
never addressed
in the literature hitherto. Since the triaxial case
is somewhat in-between the oblate and prolate cases, it is
not surprising that the diagrammes have similar form. What $\bf {is}$ 
surprising, is that, once again, all the curves seem to pass 
zero in the same point, $\;\Phi_o\;=\;\it{arccos}\,(1/\sqrt{3})\;$
(and, for symmetry reasons, in $\;\Phi_o\;=\;\pi\;-\;\it{arccos}\,(1/\sqrt{3})
\;$).

Interestingly, the Gold alignment of thermal dust fails at
the same values of $\;\Phi\;$ (Dolginov \& Mytrofanov 1976). 

This maddening coincidence makes us suppose that this special
value, $\;\Phi_o\;=\;\it{arccos}\,(1/\sqrt{3})\;$, is completely
shape-invariant and is independent from the suprathermality degree. 
So here comes
{{ {\underline{``the $\it{arccos}\,(1/\sqrt{3})-$hypothesis'':}}
No mechanical alignment of arbitrarily-shaped (not necessarily
ellipsoidal) thermal or suprathermal grains takes place, when the
magnetic line and the gas drift make angle $\;\,\it{arccos}\,(1/\sqrt{3})\;\,$
or $\;\,\pi\,-\,\it{arccos}\,(1/\sqrt{3})$}}.

\section{Conclusions}

In the article thus far, we have investigated the cross-section mechanism of
suparthermal-grain alignment in a supersonic interstellar gas stream. While the
preceding efforts had been aimed at the cases of oblate and prolate ellipsoidal
grains, 
in the current paper we studied the case of triaxial ellipsoid. We provided a 
comprehensive semianalytical-seminumerical treatment that reveals the 
dependence of the alignment measure (Rayleigh reduction factor $R$) upon the 
semiaxes' ratios and
upon the angle $\Phi$ between the magnetic line and gas drift. We provided a 
qualitative physical explanation of some of the obtained results. 

However, the most intriguing result poses a puzzle and still lacks a simple
physical explanation. This is the remarkable shape-independence of the critical
value of $\Phi$, at which $R$ vanishes and the cross-section mechanism fails. 
For all studied shapes (prolate, oblate, and triaxial with various ratios of semiaxes), this critical value
is $\;\Phi_o\;=\;\it{arccos}\,(1/\sqrt{3})\;$. We hypothesise that this special
nature of the said value of $\Phi$ is shape-independent.

\pagebreak


\appendix
\section{Relations between the observer-frame and body-frame
extinction cross sections}

Our goal is to compute the extinction cross-sections 
$\;C^{o}_{\it x}\;$ and $\;C^{o}_{\it y}\;$ for the two linear
polarisations orthogonal to one another and to the line of sight, 
$\; z_o$. These, so-called ``observer-frame'' cross-sections should be
expressed through extinction cross-sections $\;C_X\;$, $\;C_Y\;$, 
$\;C_Z\;$ appropriate to polarisations along the principal axes
$\;X,\,Y,\,Z\;$ of the granule (with $\;X\;$, $\;Y\;$, and $\;Z\;$
standing for the minimal-, middle- and maximal-inertia axes, accordingly).

As an intermediate step, let us first calculate the intensities  
$\;{E^{o}_{\it x}}^{\;2}\;$ and $\;{E^{o}_{\it y}}^{\;2}\;$
appropriate to the linear polarisations, $\;x_o\;$ and 
$\;y_o\;$. These observer-frame intensities should be
expressed through the body-frame intensities $\;E_X^2\;$, $\;E_Y^2\;$, 
$\;E_Z^2\;$ appropriate to polarisations along the principal axes. To that 
end, one has to perform a sequence of coordinate transformations. 

The first step is to express $\;E^{o}_{\it x}\;$ and 
$\;E^{o}_{\it y}\;$ through the components $\;E^{B}_{\it {x,y,z}}\;$
of $\;\bf E\, $ in the coordinate system associated with the magnetic
line. In the observer's frame, axis $\;z_o\;$ points toward the
telescope, while $\;y_o\;$ may be chosen to belong to the plane defined
by $\;z_o\;$ and $\;\bf B\;$. Then the magnetic line will lay in 
the $\;(y_oz_o)\,$ plane. A coordinate system associated with the 
magnetic field (Fig.1) may be defined with $\;y_B\;$ pointing along 
$\;\bf B\;$, and with the $\,x\,$ axis remaining untouched: $\;x_B\;
=\;x_o\,$. The angle between $\;\bf B\;$ and $\;y_o\;$ (equal to that 
between $\;z_B\;$ and $\;z_o\;$) will be called $\;\gamma\;$. Hence,
\be
E^{o}_{\it x}\;=\;E^{B}_{\it x}\;\;\;,\;\;\;\;\;\;E^{o}_{\it
y}\;=\;E^{B}_{\it y}\;\cos \gamma \;+\;E^{B}_{\it z}\;\sin \gamma \;\;.
\label{A1}
\ee
The next transformation is from $\;(x_B\, , \;y_B\, , \;z_B)\,$ to
$\;(x_J\, , \;y_J\, , \;z_J)\,$, the latter frame being affiliated
to the (precessing about $\,\bf B\,$) angular momentum $\,\bf J\,$ of
the grain. We choose $\;y_J\, $ to point along $\,\bf J\,$, at the
angular separation $\;\beta\;$ from $\;y_B\;\,$ (Fig.2). Vector $\,\bf J\,$ 
describes, about $\,\bf B\,$, a precession cone of half-angle
$\;\beta\;$; and so does $\;y_J\,$ about $\;y_B\,$. An instantaneous
position of the rotating frame $\;(x_J\, , \;y_J\, , \;z_J)\,$, with
respect to the inertial one, $\;(x_B\, , \;y_B\, , \;z_B)\,$, is
determined by angle $\;\phi\;$. We see that a transition from
$\;(x_B\, , \;y_B\, , \;z_B)\,$ is two-step: first, we must revolve
the basis about the $\;y_B$ axis by $\;\phi\;$. This will map 
axes $\;x_B\,$ and $\;z_B\,$ onto axes $\;x_J\,$ and $\;z'\,$, accordingly.
Then we rotate the basis about  $\;x_J\,$ by angle $\;\beta\;$. In the
course of a precession cycle of $\;\bf J\;$ about $\;\bf B\;$, angle 
$\;\beta\;$ remains unaltered, while $\;\phi\;$ describes the full
circle. The relations between the unit vectors are straightforward:
\be
\hat{{\bf z}}'\; =\; \hat{{\bf z}}^{B}\; \cos \phi \;- \;\hat{{\bf
x}}^{B} \; \sin \phi
\label{A2}
\ee
\be
\hat{{\bf z}}^{J}\; =\; \hat{{\bf z}}' \;\cos \beta - \hat{{\bf
y}}^{B}\; \sin \beta \;= \;
\hat{{\bf z}}^{B}\; \cos \phi \; \cos \beta \;- \;\hat{{\bf x}}^{B}
\; \sin \phi\; \cos \beta \;-\; \hat{{\bf y}}^{B}\; \sin \beta 
\label{A3}
\ee
\be
\hat{{\bf y}}^{J}\; =\; \hat{{\bf y}}^{B}\;\cos \beta\; +\; \hat{{\bf 
z}}' \;\sin \beta\;=\;
\hat{{\bf y}}^{B}\;\cos \beta\; +\; \hat{{\bf z}}^{B}\; \cos \phi
\;\sin \beta\;- \;\hat{{\bf x}}^{B}\;\sin \phi \;\sin \beta
\label{A4}
\ee
\be
\hat{{\bf x}}^{J}\; =\;\hat{{\bf x}}^{B}\; \cos \phi \;+\; \hat{{\bf z}}^{B}\;\sin \phi \;\;\;.\;\;
\label{A5}
\ee
Plugging thereof into the right-hand side of the trivial identity
$\;E^{B}_{\it x}\,\hat{{\bf{x}}}^{B}\,+\,E^{B}_{\it
y}\,\hat{{\bf{y}}}^{B}\,+\,E^{B}_{\it z}\,\hat{{\bf{z}}}^{B}\,=\,{\bf
E}\,=\,E^{J}_{\it x}\,\hat{{\bf{x}}}^{J}\,+\,E^{J}_{\it
y}\,\hat{{\bf{y}}}^{J}\,+\,E^{J}_{\it z}\,\hat{{\bf{z}}}^{J}
\;$ yields:
\ba
\;\;\;\;\;\;\;\;\;\;\;\;\;\;\;\;\;\;\;\;\;\;\;\;\;\;\;\;E^{B}_{\it x}\,=\;E^{J}_{\it x}\,\cos \phi\;-\;E^{J}_{\it y}\;\sin
\phi \;\sin \beta\;-\;E^{J}_{\it z}\,\sin \phi \;\cos \beta
\label{A6}
\ea
\ba
\;\;\;\;\;\;\;\;\;\;\;\;\;\;\;\;\;\;\;E^{B}_{\it y}\,=\;\;\;\;\;\;\;\;\;\;\;\;\;\;\;\;\;\;\;\;E^{J}_{\it y}\;\cos \beta\;\;\;\;\;\;\;\;\;\;-\;E^{J}_{\it z}\,\sin \beta
\label{A7}
\ea
\ba
\;\;\;\;\;\;\;\;\;\;\;\;\;\;\;\;\;\;\;\;\;\;\;\;\;\;\;E^{B}_{\it z}\,=\;E^{J}_{\it x}\,\sin \phi\;+\;E^{J}_{\it y}\;\cos
\phi\;\sin \beta\;+\;E^{J}_{\it z}\,\cos \phi \;\cos \beta
\label{A8}
\ea
Lastly, we take into account the position of the grain itself, relative
to coordinates $\;(x^J, \,y^J, \,z^J)\;$. As explained in the
Introduction (footnote 1), it is legitimate in the suprathermal case
to assume that the major-inertia axis $\,Z\,$ of the particle is
aligned with the angular momentum, i.e., with axis $\,y^J\;$ (for
details see Lazarian \& Efroimsky 1999). Rotation of the granule
is, thus, assumed principal and may be parametrised by the angle $\;\psi\;$
between the least-inertia axis $\;X\;$ and $\;z^J\;$ (this angle is
equal to that between the middle-inertia axis $\;Y\;$ and $\;x^J\;$). 
This gives:
\be
E^{J}_{\it x}\;=\;-\;E_X\;\sin \psi\;+\;E_Y\;\cos \psi
\label{A9}
\ee
\be
\;E^{J}_{\it y}\;=\;\;\;\;E_Z\;\;\;\;\;\;\;\;\;\;\;\;\;\;\;\;\;\;\;\;\;\;\;\;\;\;\;\;\;\;
\label{A10}
\ee
\be
\;E^{J}_{\it z}\;=\;\;\;E_X\;\cos \psi\;+\;E_Y\;\sin \psi \;\;.
\label{A11}
\ee
Combination of all the afore presented transformations results in:
\ba
\nonumber
E^{o}_{\it x}\;=\;E^{B}_{\it x}\;=\;E^{J}_{\it x}\,\cos \phi\;-\;E^{J}_{\it y}\;\sin
\phi \;\sin \beta\;-\;E^{J}_{\it z}\,\sin \phi \;\cos \beta\;=\\
=\;E_X\;(\,-\;\sin \psi \; \cos \phi\;-\;\cos \psi\;\sin \phi \;\cos
\beta) \;\;\;\;\;\;\;\;\;\;\;\;\;\;\;\;\;\;\;\;\;\;\;\;\;\;\\
\label{A12}
\nonumber
+\;E_Y\;(\cos \psi\;\cos \phi \;-\; \sin \psi\;\sin \phi \;
\cos \beta )\;-\;E_Z\;\sin \phi\;\sin \beta\;\;\;\,
\ea
and
\ba
\nonumber
E^{o}_{\it y}\,=\,E^{B}_{\it y}\,\cos \gamma \,+\,E^{B}_{\it z}\,
\sin \gamma \,=\;\;\;\;\;\;\;\;\;\;\;\;\;\;\;\;\;\;\;\;\;\;\;\;\;\;\;\;\;\;\;\;\;\;\;\;\;\;\;\;\;\;\;\;\;\;\;\;\;\;\;\;\;\;\;\;\;\;\;\;\;\;\;\;\;\;\;\;\;\;\;\;\;\;\;\\
\nonumber
\left( E^{J}_{\it y}\,\cos \beta\,-\,E^{J}_{\it z}\,
\sin \beta \right) \,\cos \gamma \,+\,\left(E^{J}_{\it x}\,\sin
\phi\,+\,E^{J}_{\it y}\,\cos \phi\,\sin \beta \;+\;E^{J}_{\it z}\,\cos \phi \;\cos \beta\right) \,\sin \gamma\;\;\;\;\;\;
\ea
\ba
\nonumber
=\,E_X\;(\,-\;\sin \psi\; \sin \gamma\;\sin \phi\;-\;\cos \psi\;\cos
\gamma\;\sin \beta\;+\;\cos \psi\;\cos \phi\;\sin \gamma \; \cos
\beta) \\
+\;E_Y\;(\cos \psi\; \sin \gamma\;\sin \phi\;-\;\sin \psi\;\cos
\gamma\;\sin \beta\;+\;\sin \psi\;\cos \phi\;\sin \gamma \; \cos
\beta)\;\;\;\;\;\;\\
\nonumber
+\;E_Z\;(\cos \gamma\;\cos \beta\;+\;\sin \gamma\;\cos
\phi\;\sin \beta)\;\;\;,\;\;\;\;\;\;\;\;\;\;\;\;\;\;\;\;\;\;\;\;\;\;\;\;\;\;
\;\;\;\;\;\;\;\;\;\;\;\;\;\;\;\;\;\;\;\;\;\;\;\;\;
\ea
whence the averaged-over-the-ensemble intensities read:
\ba
\nonumber
\langle {E_x^o}^2 \rangle\;=\;\frac{1}{4}\;E_X^2\;\left(1\;+\;{\Big{\langle}}
\cos^2 \beta
{\Big{\rangle}}\right)\;+\;\frac{1}{4}\;E_Y^2\;\left(1\;+\;
{\Big{\langle}}
\cos^2 \beta
{\Big{\rangle}}\right)\;+\;\frac{1}{2}\;E_Z^2\;{\Big{\langle}} 
\sin^2 \beta {\Big{\rangle}}\;=\\
\nonumber\\
\nonumber
\frac{1}{3}\;\left(E_X^2\;+\;E_Y^2\;+\;E_Z^2\right)\;+\;\frac{1}{3}\;
\left[ \frac{1}{2}\;\left(E_X^2\;+\;E_Y^2 \right)\;-\;E_Z^2 \right]\;
\frac{3}{2}\;\left({\Big{\langle}} \cos^2 \beta {\Big{\rangle}}\;-\;
\frac{1}{3}\right)
\ea
\ba
=\;\frac{1}{3}\;\left(E_X^2\;+\;E_Y^2\;+\;E_Z^2\right)\;+\;\frac{1}{3}\;
\left[ \frac{1}{2}\;\left(E_X^2\;+\;E_Y^2 \right)\;-\;E_Z^2 \right]\;R
\label{A14}
\ea
and
\ba
\nonumber
{\Big{\langle}} {E_y^o}^2 {\Big{\rangle}}\; = \; E_X^2\; \left(
\frac{1}{4}\;\sin^2 \gamma\;+\;\frac{1}{2}\;\cos^2 \gamma\;\,{\Big{\langle}}
\sin^2 \beta {\Big{\rangle}} \;  +\;\frac{1}{4}\;\sin^2 \gamma\;\,
{\Big{\langle}} \cos^2 \beta {\Big{\rangle}}  \right)\;+
\ea
\ba
\nonumber
\;\;\;\;\;\;\;\;\;\;\;E_Y^2\;\left( \frac{1}{4}\,\sin^2 \gamma\;+\;\frac{1}{2}\,\cos^2 \gamma\;{\Big{\langle}}
\sin^2 \beta {\Big{\rangle}}\;+\;\frac{1}{4}\,\sin^2 \gamma\;{\Big{\langle}}
\cos^2 \beta {\Big{\rangle}}\right)\;+
\ea
\ba
\nonumber
E_Z^2\;\left(\cos^2 \gamma\;\,{\Big{\langle}} \cos^2 \beta {\Big{\rangle}}
\;+\;\frac{1}{2}\;\sin^2 \gamma\;\,{\Big{\langle}} \sin^2 \beta {\Big{\rangle}}
\right)\;=\;\;\;\;\;\;\;\;\;\,
\ea
\ba
\nonumber
\frac{1}{3}\;\left(E_X^2\;+\;E_Y^2\;+\;E_Z^2\right)\;+\;\left(\frac{1}{3}\;-\;\cos^2
\gamma \right)\; \left[ \frac{1}{2}\;\left(E_X^2\;+\;E_Y^2 \right)\;-\;E_Z^2 \right]\;
\frac{3}{2}\;\left(\langle \cos^2 \beta \rangle\;-\;\frac{1}{3}\right)
\ea
\ba
=\;\frac{1}{3}\;\left(E_X^2\;+\;E_Y^2\;+\;E_Z^2\right)\;+\;\left(\frac{1}{3}\;-\;\cos^2
\gamma \right)\; \left[ \frac{1}{2}\;\left(E_X^2\;+\;E_Y^2 \right)\;-\;E_Z^2 \right]\;R
\label{A15}
\ea
Following the established tradition, we single out the
so-called Rayleigh reduction factor
\be
R\;=\;\frac{3}{2}\;\left( \; {\Big{\langle}} \; \cos^2 \beta \; {\Big{\rangle}}
\; - \; \frac{1}{3} \; \right)\;\;.
\label{A16}
\ee
In (\ref{A14}) - (\ref{A15}) averaging over $\;\psi\;$ and $\;\phi\;$ 
implies simply $\;(2\pi)^{-2}\,\int_{0}^{2\pi}d\psi\;\int_{0}^{2\pi}d\phi\;$, 
while the averaging over $\;\beta\;$ remains so far unspecified; the 
appropriate distribution depends upon the physics of gas-grain interaction 
and is calculated in Section V.

Now, that we have expressed the observable intensities $\;\langle {E_x^o}^2 
\rangle\;$ and $\;\langle {E_y^o}^2 \rangle\;$ via those 
appropriate to polarisations along the body axes, we can write down
similar expressions interconnecting the extinction cross
sections. Since the extinction is merely attenuation of power from the
incident beam, the contributions to cross-section from the body-axes' 
directions will be proportional to the mean shares of power
appropriate to these three axes:
\be
C^o_x\;=\;\frac{C_X \;+\;C_Y\;+\;C_Z}{3}\;+\;\frac{1}{3}\;
\left[ \; \frac{1}{2}\;\left(C_X\;+\;C_Y \right)\;-\;C_Z \; \right]\;R
\label{A17}
\ee
and 
\be
C^o_y\;=\;\frac{C_X \;+\;C_Y\;+\;C_Z}{3}\;+\;\left(\frac{1}{3}\;-\;\cos^2
\gamma \right)\; \left[\;\frac{1}{2}\;\left(C_X\;+\;C_Y
\right)\;-\;C_Z\; \right]\;R
\label{A18}
\ee
wherefrom 
\be
C^o_x\;-\;C^o_y\;=\;\left[\;\frac{1}{2}\;\left(C_X\;+\;C_Y\right)\;-\;C_Z \;
\right]\;R\;\cos^2 \gamma\;\;\;.
\label{A19}
\ee

\section{Ellipsoidal granule placed in a magnetic field and gas flow. Several
useful formulae.}

Here we prove several formulae used in the text. Let us begin with 
relations for the angle $\;\alpha\;$ between the gas-flow direction 
$\;\bf f\;$ and the maximum-inertia axis $\;Z\;$ of the dust particle
(Fig. 4). Dot product of the appropriate unit vectors
\be
\hat {\bf y}_J\;=\;{\hat {\bf y}}_B\;\cos \beta\;+\;\hat {\bf z}_B \;\sin 
\beta\;\cos \phi\;+\;\hat {\bf x}_B\;\sin \beta\;\sin \phi
\label{B1}
\ee
and
\be
{\bf f}\;=\;{\hat {\bf y}}_B\;\cos \Phi\;+\;\hat {\bf z}_B \; \sin \Phi
\label{B2}
\ee
leads to formula:
\be
\cos \alpha\;=\;\cos \Phi\;\cos \beta\;+\;\sin \Phi\;\sin \beta\;\sin \phi\;\;.
\label{B3}
\ee
Another important relation is evident from Fig. 3:
\be
\cos \alpha\;=\;\frac{Z_f}{\sqrt{X_f^2\;+\;Y_f^2\;+\;Z_f^2}} \;\;.
\label{B4}
\ee
On Fig.3, the projection of $\;\it \bf f\;$ on 
$\;(X,\,Y)\;$ will make an angle $\,\eta\,$ with axis $\,X\,$, such that
\be
\cos \eta\;\sin\alpha\;=\;\frac{X_f}{\sqrt{X_f^2\;+\;Y_f^2\;+\;Z_f^2}} \;\;\;
,\;\;\;\;\;\;\;\; \sin \eta\;\sin\alpha\;=\;\frac{Y_f}{\sqrt{X_f^2\;+\;Y_f^2\;
+\;Z_f^2}} 
\label{B5}
\ee
Formulae (\ref{B4}) and (\ref{B5}) will enable us to calculate the grain's
cross section relative to the wind. The lines of gas flow, which are 
tangential to the ellipsoid surface, touch this surface in points that 
altogether constitute a curve. It is the ellipse hatched on Fig. 5. Its area is:
\be
\Sigma\;=\;\pi\;|{\robold}_{min}\;\times\;{\robold}_{max}|\;=\;\pi\;{\rho}_{min}\;{\rho}_{max}
\label{B6}
\ee
${\robold}_{min}\;$ and $\;{\robold}_{max}\;$ being its semiaxes.
Projection of $\;\Sigma\;$ toward the plane perpendicular 
to the gas flow is the cross section $\;S\;$ of the granule, as seen by the
observer looking at it along the line of wind (Fig. 6). In other words, $\;S\;$
is the ``shadow'' that the granule casts. Evidently,
\be
S\;=\;|\cos \theta|\;\,\Sigma
\label{B7}
\ee
$\theta\;$ being the angle between the vector $\;\bf f\;$ of the gas flow
and vector 
\be
{\bf n}\;\equiv\;  {\robold}_{min}\;\times\;{\robold}_{max}
\label{B8}
\ee
orthogonal to the hatched ellipse. As follows from (\ref{B4} - \ref{B5}),
\ba
\nonumber
|\cos \theta|\;=\;\frac{{\bf n\;\cdot f}}{|\bf n|\;|\bf f|}\;=\;
\frac{|n_x\;X_f\;+\;n_y\;Y_f\;+\;n_z\;Z_f|}{\left(\Sigma/\pi\right)\;\sqrt{X_f^2\,+\,Y_f^2\,+\,Z_f^2}}\;=\\
\nonumber\\
\frac{\pi}{\Sigma}\;\;|n_x\;\cos \eta\;\sin \alpha \;+\;n_y\;\sin
\eta\;\sin \alpha\;+\;n_z\;\cos \alpha\,|\;.
\label{B9}
\ea
Combining the above with (\ref{B7}), we arrive to:
\be
S\;=\;\pi\;\,|n_x\;\cos \eta\;\sin \alpha \;+\;n_y\;\sin
\eta\;\sin \alpha\;+\;n_z\;\cos \alpha\,|\;\;.
\label{B10}
\ee

\section{Cross Section of a granule arbitrarily oriented in
the gas stream}

Every point $\;(X,\,Y,\,Z)\;$ belonging to the surface
of the ellipsoidal grain obeys equation
\be
g(X,\,Y,\,Z;\,a,\,b,\,c)\;\equiv\;\frac{X^2}{a^2}\;+\;\frac{Y^2}{b^2}\;+\;
\frac{Z^2}{c^2}\;=\;1\;\;.
\label{C1}
\ee
The points, that constitute the boundary of the hatched figure $\,\Sigma\,$
on Fig.5, obey (\ref{C1}), along with one more relation. That second
one is the condition of flow being tangential to the surface
in these points. Stated alternatively: a normal to the ellipsoid
in point $\;(X,\,Y,\,Z)\;$ is given by vector $\;(X/a^2,\,Y/b^2,\,Z/c^2)\;$, 
and the flow must be orthogonal to this normal:
\be
h(X,\,Y,\,Z;\,X_f,\,Y_f,\,Z_f)\;\equiv\;\frac{X\;X_f}{a^2}\;+\;
\frac{Y\;Y_f}{b^2}\;+\;\frac{Z\;Z_f}{c^2}\;=\;0\;\;.
\label{C2}
\ee
To find vectors $\;{\robold}_{min}\;$ and $\;{\robold}_{max}\;$  pointing from 
the centre to the closest and the farthest points of the ellipse $\;\Sigma\;$,
one has to employ the variational method:
\be
\frac{\partial}{\partial X,\,Y,\,Z}\;\left( {\it l}^2\;-\;\lambda\;g\;-
\;\mu\;h \right)\;=\;0
\label{C3}
\ee
where $\;{\it l}^2\;\equiv\;X^2\;+\;Y^2\;+\;Z^2\;$, and $\;\lambda,\,\mu\;$ are 
Lagrange multipliers. The latter equation gives us the values of $\;X,\,Y,\,Z
\;$ appropriate to the extremal distances from the origin, in assumption that 
constraints (\ref{C1}) and (\ref{C2}) hold. Three equations
(\ref{C3}), for  $\;X,\,Y\,$ and $\,Z\;$, yield:
\be
X\;=\;\frac{\mu \;X_f}{2\;\left(a^2\;-\;\lambda  \right)}\;\;,\;\;\;
Y\;=\;\frac{\mu \;Y_f}{2\;\left(b^2\;-\;\lambda  \right)}\;\;,\;\;\;
Z\;=\;\frac{\mu \;Z_f}{2\;\left(c^2\;-\;\lambda  \right)}
\label{C4}
\ee
for the extremal points where vectors $\;\pm\,{\robold}_{min}\;$ and
$\;\pm\,{\robold}_{max}\;$ end. To find the values of $\;\lambda\;$,
plug (\ref{C4}) in (\ref{C2}):
\be
h\;=\;\mu\;\frac{X^2_f}{2\;a^2\;\left(a^2\;-\;\lambda \right)}\;+\;
      \mu\;\frac{Y^2_f}{2\;b^2\;\left(b^2\;-\;\lambda \right)}\;+\;
      \mu\;\frac{Z^2_f}{2\;c^2\;\left(c^2\;-\;\lambda \right)}\;=\;0\;\;.
\label{C5}
\ee
This entails:
\be
\lambda_{1,2}\;=\;\frac{-\;Q\;\pm\;\sqrt{Q^2\;-\;4\;P\;R}}{2\;P}
\label{C6}
\ee
where
\be
P\;\equiv\;\left(\frac{X_f}{a}\right)^2\;+\;\left(\frac{Y_f}{b}\right)^2\;+\;
\left(\frac{Z_f}{c}\right)^2\;\;,\;\;\;\;
\label{C7}
\ee
\be
Q\;\equiv\;-\;\left[\;
\left(\frac{X_f}{a}\right)^2\;\left(b^2\;+\;c^2\right)\;+\;
\left(\frac{Y_f}{b}\right)^2\;\left(c^2\;+\;a^2\right)\;+\;
\left(\frac{Z_f}{c}\right)^2\;\left(a^2\;+\;b^2\right)\;
\right]\;\;,\;\;\;\;
\label{C8}
\ee
\be
R\;\equiv\;\left(\frac{X_f}{a}\right)^2\;b^2\;c^2\;+\;
           \left(\frac{Y_f}{b}\right)^2\;c^2\;a^2\;+\;
           \left(\frac{Z_f}{c}\right)^2\;a^2\;b^2\;\;\;.
\label{C9}
\ee
Substitution of (\ref{C4}) in (\ref{C1}) will lead to the expression
for $\;\mu\;$:
\be
\mu^2\;=\;4\;\left(\left(\frac{X_f}{a\;\left(a^2\,-\,\lambda\right)}\right)^2\;
 +\;               \left(\frac{Y_f}{b\;\left(b^2\,-\,\lambda\right)}\right)^2\;
 +\;               \left(\frac{Z_f}{c\;\left(c^2\,-\,\lambda\right)}\right)^2\;
  \right)^{-1}
\label{C10}
\ee
Since we have two acceptable values for $\;\lambda\;$, we shall obtain four
different values for $\;\mu\;$:
\ba
\mu_1\;=\;2\;\left( \left(\frac{X_f}{a\;\left(a^2\,-\,\lambda_1\right)}\right)^2\;+\;
                   \left(\frac{Y_f}{b\;\left(b^2\,-\,\lambda_1\right)}\right)^2\;+\;
                   \left(\frac{Z_f}{c\;\left(c^2\,-\,\lambda_1\right)}\right)^2  \right)^{-1/2}\;\;,\;\;\;\;
\label{C11}
\ea
\ba
\mu_2\;=\;2\;\left(\left(\frac{X_f}{a\;\left(a^2\,-\,\lambda_2\right)}\right)^2\;+\;
                   \left(\frac{Y_f}{b\;\left(b^2\,-\,\lambda_2\right)}\right)^2\;+\;
                   \left(\frac{Z_f}{c\;\left(c^2\,-\,\lambda_2\right)}\right)^2   \right)^{-1/2}\;\;,
\label{C12}
\ea
and $\;\mu_3\;=\;-\;\mu_1\;,\;\;\mu_4\;=\;-\;\mu_2\;$.
Simply from looking at how $\;\lambda\;$ and  $\;\mu\;$ enter expressions 
(\ref{C4}) for extremal-point coordinates, we see that a change of sign of 
$\;\mu\;$ (with $\;\lambda\;$ fixed) corresponds merely to a switch from
$\;{\robold}_{min}\;$ (or $\;{\robold}_{max}\;$) to $\;-\,{\robold}_{min}\;$ 
(or $\;-\,{\robold}_{max}\;$, appropriately). Since it is irrelevant, for our 
purposes, which of the two opposite farmost from the origin points to call
$\;{\robold}_{max}\;$ and which to call $\;-\,{\robold}_{max}\;$ (and,
similarly, which of the two closemost to the origin points to call 
$\;{\robold}_{min}\;$ and which to call $\;-\,{\robold}_{max}\;$), we shall
take the positive values of $\;\mu\;$ only. As a result, a 
substitution of $\;\lambda_1\;$ and  $\;\mu_1\;$ in (\ref{C4}) will give us the
coordinates of one of the two farthest, from the centre, points of the
boundary of the hatched ellipse $\;\Sigma\;$.  
Similarly, plugging of $\;\lambda_2\;$ and $\;\mu_2\;$ will provide us with the
coordinates of one of the two closest points. The chosen farthest and closest
points will have coordinates $\;(\rho_{max})_{x,y,z}\;$ and 
$\;(\rho_{min})_{x,y,z}\;\,$, appropriately:
\be
(\rho_{max})_{x}\;=\;\frac{\mu_1\;X_f}{2\;\left(a^2\,-\,\lambda_1\right)}\;\;,
\;\;\;
(\rho_{max})_{y}\;=\;\frac{\mu_1\;Y_f}{2\;\left(b^2\,-\,\lambda_1\right)}\;\;,
\;\;\;
(\rho_{max})_{z}\;=\;\frac{\mu_1\;Z_f}{2\;\left(c^2\,-\,\lambda_1\right)}\;\;,
\label{C13}
\ee
and
\be
(\rho_{min})_{x}\;=\;\frac{\mu_2\;X_f}{2\;\left(a^2\,-\,\lambda_2\right)}\;\;,
\;\;\;
(\rho_{min})_{y}\;=\;\frac{\mu_2\;Y_f}{2\;\left(b^2\,-\,\lambda_2\right)}\;\;,
\;\;\;
(\rho_{min})_{z}\;=\;\frac{\mu_2\;Z_f}{2\;\left(c^2\,-\,\lambda_2\right)}\;\;.
\label{C14}
\ee
Further substitution of these expressions in (\ref{6.4}) and in its
cyclic transpositions will lead to:
\be
n_x\;X_f\;=\;X_f\;Y_f\;Z_f\;\frac{\mu_1\;\mu_2}{4}\;\left[
\frac{1}{\left(b^2\,-\,\lambda_2 \right)\,\left(c^2\,-\,\lambda_1 \right)}\;-\;\frac{1}{\left(c^2\,-\,\lambda_2\right)\,\left(b^2\,-\,\lambda_1 \right)}
\right]
\label{C15}
\ee
\be
n_y\;Y_f\;=\;X_f\;Y_f\;Z_f\;\frac{\mu_1\;\mu_2}{4}\;\left[
\frac{1}{\left(c^2\,-\,\lambda_2  \right)\,\left(a^2\,-\,\lambda_1 \right)}\;-\;\frac{1}{\left(a^2\,-\,\lambda_2   \right)\,\left(c^2\,-\,\lambda_1   \right)}
\right]
\label{C16}
\ee
\be
n_z\;Z_f\;=\;X_f\;Y_f\;Z_f\;\frac{\mu_1\;\mu_2}{4}\;\left[
\frac{1}{\left(a^2\,-\,\lambda_2 \right)\,\left(b^2\,-\,\lambda_1\right)}\;-\;\frac{1}{\left(b^2\,-\,\lambda_2 \right)\,\left(a^2\,-\,\lambda_1\right)}
\right]
\label{C17}
\ee
Thence the cross section $\;S\;$ of the grain, relative to the gas-flow
direction, will read:
\ba
\nonumber
S\;=\;\frac{\pi}{4}\;\mu_1\;\mu_2\;{|X_f\;Y_f\;Z_f|}
\;\left[
\frac{1}{\left(b^2\,-\,\lambda_2
\right)\,\left(c^2\,-\,\lambda_1\right)}\;-\;
\frac{1}{\left(c^2\,-\,\lambda_2 \right)\,\left(b^2\,-\,\lambda_1
\right)}\right.\;\;\;\\
\;+\;
\frac{1}{\left(c^2\,-\,\lambda_2 \right)\,\left(a^2\,-\,\lambda_1
\right)}\;-\;\frac{1}{\left(a^2\,-\,\lambda_2
\right)\,\left(c^2\,-\,\lambda_1 \right)}\;\;\;\\
\nonumber
\;\;\;\;\;\;\;\;\;\;\;\;\;\;\;\;\;\;\;\;\left.\;+\;
\frac{1}{\left(a^2\,-\,\lambda_2 \right)\,\left(b^2\,-\,\lambda_1\right)}\;-\;\frac{1}{\left(b^2\,-\,\lambda_2 \right)\,\left(a^2\,-\,\lambda_1\right)}
\;\right]
\label{C18}
\ea
where $\;\lambda_{1,2}\;$ and $\;\mu_{1,2}\;$ are functions of 
$\;(X_f,\,Y_f,\,Z_f)\;$, the latter being functions of
angles $\;\eta\;$ and $\;\alpha\;$ (while $\;\alpha\;$ is, in its
turn, depends upon $\;\beta\;$, $\;\Phi\;$, and $\;\phi\;$). All in
all, the ``shadow'' area $\;S\;$ turns to be a function of angles 
$\;\beta\;$, $\;\Phi\;$, $\;\eta\;$, and $\;\phi\;$.

\end{document}